\documentclass[12pt,a4paper]{article}
\usepackage{a4wide}
\usepackage{latexsym}
\usepackage{epsf}
\usepackage{amssymb}
\makeatletter
\@addtoreset{equation}{section}
\makeatother

\begin{document}

\begin{flushright}
\small
IFT-UAM/CSIC-05-26\\
{\bf hep-th/0506056}\\
June $7$th, $2005$
\normalsize
\end{flushright}

\begin{center}

\vspace{2cm}

{\Large {\bf All the supersymmetric configurations\\[.6cm] 
of $N=4,d=4$ supergravity}}

\vspace{2.5cm}

{\bf\large Jorge Bellor\'{\i}n}\footnote{E-mail: {\tt Jorge.Bellorin@uam.es}},
{\bf\large and Tom{\'a}s Ort\'{\i}n}\footnote{E-mail: {\tt Tomas.Ortin@cern.ch}}

\vspace{1cm}

{\it Instituto de F\'{\i}sica Te\'orica UAM/CSIC\\
  Facultad de Ciencias C-XVI,
  C.U. Cantoblanco,  E-28049-Madrid, Spain}\\

\vspace{3cm}


{\bf Abstract}

\end{center}

\begin{quotation}

\small

All the supersymmetric configurations of pure, ungauged, $N=4,d=4$
supergravity are classified in a formalism that keeps manifest the S and T
dualities of the theory. We also find simple equations that need to be
satisfied by the configurations to be classical solutions of the theory.
While the solutions associated to null Killing vectors were essentially
classified by Tod (a classification that we refine), we find new
configurations and solutions associated to timelike Killing vectors that do
not satisfy Tod's rigidity hypothesis (hence, they have a non-trivial $U(1)$
connection) and whose supersymmetry projector is associated to 1-dimensional
objects (strings), although they have a trivial axion field.

\end{quotation}

\newpage

\pagestyle{plain}


\tableofcontents

\newpage

\section{Introduction and main results}

Classical supersymmetric solutions of supergravity theories have played a very
important role in many advances in string theory for the past 15 years and are
still the subject of much interest since they include, for example,
backgrounds (possibly with branes and fluxes) for string model-building and
supersymmetric objects such as black holes, supertubes and, as it has been
discovered recently in Ref.~\cite{Elvang:2004rt}, black rings.

It is, thus, a very interesting problem to try to find or at least classify
and characterize the supersymmetric solutions of (ideally all) supergravity
theories. There have been many interesting results in the literature on this
program, starting with the work of Gibbons and Hull in $N=2,d=4$ supergravity
\cite{Gibbons:1982fy}, completed in the seminal paper Ref.~\cite{Tod:1983pm}
by Tod, who, starting from the Killing spinor equations (KSEs) of that theory
and using all the integrability conditions and properties derived from them,
assuming the existence on one Killing spinor, was able to find, for the first
time, all the field configurations (metric and vector field strength) for
which the KSEs could be solved. His classification included field
configurations which may or may not satisfy the classical equations of motion.

It was only 12 years later that a similar task was undertaken again by Tod,
who in Ref.~\cite{Tod:1995jf} studied the supersymmetric solutions of pure,
ungauged, $N=4,d=4$ supergravity, achieving a complete classification of the
degenerate case (in which the Killing spinor gives rise to a null Killing
vector) and only a partial classification of the non-degenerate case (in which
the Killing spinor gives rise to a timelike Killing vector), since he had to
assume a hypothesis of internal rigidity that he could not prove. The
internally rigid cases were very interesting, though, since, as shown in
Ref.~\cite{Bergshoeff:1996gg} they included all known the supersymmetric
black-hole solutions of the theory, constructed by different methods and
studied in Refs.~\cite{Gibbons:1982ih}-\cite{Garcia:1995qz}. By deformation of
the supersymmetric black-hole solutions, the most general non-extremal
black-hole solutions of the theory were constructed in
Ref.~\cite{Lozano-Tellechea:1999my}.

The program enjoyed a revival when a new maximally supersymmetric solution of
$N=2B,d=10$ supergravity was discovered in Ref.~\cite{Blau:2001ne}, analogous
maximally supersymmetric solutions of 11-dimensional and $N=2,d=4$
supergravity \cite{Kowalski-Glikman:1984wv,Kowalski-Glikman:1985im} were
rediscovered and additional maximally supersymmetric solutions of the same
kind were found in 5 and 6 dimensions in Ref.~\cite{Meessen:2001vx}. The
classification of the maximally supersymmetric vacua of the 11- and
10-dimensional theories was completed in
Refs.~\cite{Figueroa-O'Farrill:2002ft,Figueroa-O'Farrill:2002xg}. It was then
realized that we still had a very incomplete knowledge of the
\textit{landscape} of supersymmetric solutions of even the simplest
supergravity theories and that new interesting supersymmetric solutions could
be found by a systematic study of the solutions of the KSEs.

This was done in Ref.~\cite{Gauntlett:2002nw} for the minimal 5-dimensional
supergravity, using a technique different from Tod's, who used the
Newman-Penrose formalism. In this work, the KSEs were translated into a set of
differential equations on all the tensors that could be constructed as
bilinears of the Killing spinors, which can be managed by more standard
techniques. Several of the new solutions found in this work have had a great
impact: a new maximally supersymmetric solution of G\"odel type and the
supersymmetric black rings
\cite{Elvang:2004rt,Gauntlett:2004wh,Gauntlett:2004qy}\footnote{More general
  black ring solutions have also been found in Refs.~\cite{Bena:2004de}} and
generalizations that lead, for instance, to supersymmetric 4-dimensional
rotating two- and one-black-hole solutions
\cite{Elvang:2005sa,Gaiotto:2005xt,Bena:2005ni}.

This work was generalized to minimal gauged 5-dimensional supergravity in
Ref.~\cite{Gauntlett:2003fk} and then analogous results were obtained for
minimal 6-dimensional supergravity in
Refs.~\cite{Gutowski:2003rg,Chamseddine:2003yy} and for $N=2,d=4$ $U(1)$
gauged supergravity in Ref.~\cite{Caldarelli:2003pb}. There is also extensive
work on the 11-dimensional and $N=2A,B,d=10$ supergravities (see
e.g.~Refs.~\cite{Gauntlett:2002fz}-\cite{Gran:2005kg}), although a complete
classification is still lacking.

In this paper we return to the problem of finding all the supersymmetric
configurations of $N=4,d=4$ supergravity, partially solved by Tod in
Ref.~\cite{Tod:1995jf}. We use tensor methods, based on the bilinears of
complex chiral spinors with $SU(4)$ indices, which allows us to keep manifest
the S and T dualities of the theory at all stages in our analysis and in the
field configurations, as it happens in the solutions studied in
Ref.~\cite{Bergshoeff:1996gg}. The formalism used here can be used as starting
point for the study of more complicated theories such as gauged and
matter-coupled $N=4,d=4$ theories and there is work in progress in these
directions.

We are going to describe our main results in a moment but, before, it is worth
explaining why $N=4,d=4$ supergravity is an interesting theory from the string
theory point of view. The toroidal compactification of the heterotic string
effective action ($N=1,d=10$ supergravity coupled to 16 vector multiplets)
gives ungauged $N=4,d=4$ supergravity coupled to 22 (matter) vector multiplets
\cite{Chamseddine:1980cp} and a consistent truncation of the matter vector
multiplets gives the pure theory that we study here. Thus, all the solutions
we will find are also solutions of the heterotic string effective action. The
truncation preserves some of the $SO(6,22;\mathbb{Z})$ T duality symmetry and
the theory is invariant under the continuous group $SO(6)\sim SU(4)$ which
naturally occurs as a \textit{hidden symmetry} of the theory\footnote{The
  first $N=4,d=4$ theory, constructed in Ref.~\cite{Das:1977uy} had only
  $SO(4)$ invariance. We will work with the $SU(4)$ theory of
  Ref.~\cite{Cremmer:1977tt}.} \cite{Cremmer:1977zt}. The theory also has an S
duality which manifest itself as a continuous $SL(2,\mathbb{R})$
\textit{hidden symmetry}. It was this symmetry which lead to the S duality
conjectures in the corresponding superstring theory
\cite{Font:1990gx}-\cite{Hull:1994ys}. We will also keep this symmetry
manifest at all stages in our analysis.

Let us now describe our results for supersymmetric solutions, leaving the more
general conditions for supersymmetric configurations which may or may not be
solutions of the equations of motion.

There are two types of supersymmetric solutions in $N=4,d=4$ supergravity
admitting at least one Killing spinor $\epsilon_{I}$, that can be
characterized by the causal nature of the vector bilinear
$V^{a}=i\bar{\epsilon}^{I}\gamma^{a}\epsilon_{I}$, which is always a
non-spacelike Killing vector.

\begin{description}
\item[Timelike $V^{a}$:] Supersymmetric solutions are determined by a choice
  of 6 time-independent complex scalars $M_{IJ}$ and a complex scalar $\tau$
  that in general may depend on the spatial coordinates $x,z,z^{*}$.  The
  $M_{IJ}$s have to satisfy two conditions:

  \begin{enumerate}
  \item Their matrix must have vanishing Pfaffian

    \begin{equation}
     \varepsilon^{IJKL}M_{IJ}M_{KL}=0\, . 
    \end{equation}
    
  \item They must be such that the 1-form $\xi$ defined in
    Eq.~(\ref{eq:xidef}) takes the form 

\begin{equation}
\xi = \pm{\textstyle\frac{i}{2}}
(\partial_{\underline{z}}Udz -\partial_{\underline{z}^{*}}Udz^{*}) 
+{\textstyle\frac{1}{2}}d\lambda \, ,
\end{equation}

for some real functions $U(z,z^{*})$ and $\lambda(x,z,z^{*})$\footnote{A
  general \textit{Ansatz} that satisfies these two conditions is given in
  Eq.~(\ref{eq:ansatzm}).}. Observe that it is the function $U$ that makes
$\xi$ non-trivial.

  \end{enumerate}
  
  $\tau$ and $M_{IJ}$ must satisfy the 3-dimensional differential equations

\begin{equation}
\nabla_{\underline{i}}(e^{2i\lambda}A^{\underline{i}}) 
-e^{2i\lambda}[\partial_{\underline{z}}(e^{-2U})A_{\underline{z}^{*}}
-\partial_{\underline{z}^{*}}(e^{-2U})A_{\underline{z}}]=0\, , 
\end{equation}


\noindent
both for 

\begin{equation}
A= \frac{d\tau}{\Im {\rm m}\tau |M|^{2}}\, ,\,\,\,\, {\rm and}\,\,\,\,
A= \frac{d[(\Im {\rm m}\tau)^{1/2}M^{IJ}]}{\Im {\rm m}\tau |M|^{2}}\, ,
\hspace{1cm}
|M|^{2}=M^{IJ}M_{IJ}\, ,
\end{equation}

\noindent 
relative to the 3-dimensional metric 

\begin{equation}
\gamma_{\underline{i}\underline{j}}dx^{\underline{i}}dx^{\underline{j}}
= dx^{2}+2e^{2U(z,z^{*})}dzdz^{*}\, ,
\end{equation}

\noindent
whose triviality is associated to that of the connection $\xi$.  Then, the
metric is given by

\begin{equation}
ds^{2} = |M|^{2}(dt+\omega)^{2} -|M|^{-2}(dx^{2}+2e^{2U}dzdz^{*})\, ,
\end{equation}

\noindent
where $\omega=\omega_{\underline{i}}dx^{i}$ satisfies

\begin{equation}
f_{ij} = 4|M|^{-2}\epsilon_{ijk}
\left(\xi_{k}-\frac{\partial_{k}\Re {\rm e}\tau}{4\Im {\rm m}\tau}\right)\, ,
\hspace{1cm}
f_{\underline{i}\underline{j}}\equiv
2\partial_{[\underline{i}}\omega_{\underline{j}]}\, .  
\end{equation}

\noindent
again relative to the above 3-dimensional metric and the vector field
strengths are given by

\begin{equation}
 F_{IJ} =  
\frac{1}{2|M|^{2}}\left\{ 
 \hat{V}\wedge dE_{IJ}
-{}^{\star}\!\!
\left[\hat{V}\wedge \left(\frac{\Re {\rm e}\,\tau}{\Im {\rm m}\,\tau}dE_{IJ} 
-\frac{1}{\Im {\rm m}\,\tau} dB_{IJ} \right)\right]
\right\}\, ,
\end{equation}

\noindent
where

\begin{equation}
  \begin{array}{rcl}
\hat{V} & = & \sqrt{2}|M|^{2}(dt +\omega)\, ,\\
& & \\
E_{IJ} & = & 2\sqrt{2} (\Im {\rm m}\,\tau)^{-1/2} (M_{IJ}+\tilde{M}_{IJ})\, ,\\
& & \\
B_{IJ} & = & 2\sqrt{2} (\Im {\rm m}\,\tau)^{-1/2} 
(\tau M_{IJ} +\tau^{*} \tilde{M}_{IJ}) \, ,\\
  \end{array}
\end{equation}

Examples of solutions corresponding to specific choices of $M_{IJ}$ and $\tau$
are given in Section~\ref{sec-susyconfigsolu}, but it is clear that there are
two different kinds of solutions which differ by the triviality of the
connection $\xi$ and the 3-dimensional metric. The case in which $\xi$ is
trivial was completely solved by Tod in Ref.~\cite{Tod:1995jf}.

\item[Null $V^{a}$] This case (called \textit{degenerate} by Tod) was
  essentially solved by Tod in Ref.~\cite{Tod:1995jf}, but we study it here
  again for the sake of completeness and to refine his results.  There are two
  subcases which we call $A$ and $B$ and which are associated to $U(1)$
  holonomy in a null direction and in a pair of spacelike directions,
  respectively, and describe $pp$-waves and the \textit{stringy cosmic
    strings} of Ref.~\cite{Greene:1989ya}.

  \begin{description}
  \item[Case A:] Each solution in this class is determined by 5 arbitrary
    functions of $u$: $\phi_{I},\tau$. Given these functions, the metric and
    vector field strengths are given by

    \begin{equation}
      \begin{array}{rcl}
ds^{2} & = & 2 du [dv + K(u,z,z^{*}) du] -2dzdz^{*}\, ,\\
& & \\
F_{IJ} & = & {\textstyle\frac{1}{2}}
(\mathcal{F}_{IJ}+{\textstyle\frac{1}{2}}
\varepsilon_{IJKL}\mathcal{F}^{KL})du\wedge dz^{*}\, ,\\
      \end{array}
    \end{equation}

\noindent
where 

\begin{equation}
  \begin{array}{rcl}
\mathcal{F}_{IJ}
& = & {\displaystyle\frac{8\sqrt{2}}{(\Im{\rm m}\, \tau)^{1/2}}}
\dot{\phi}_{[I}\phi_{J]}\, ,\\
& & \\
2\partial_{\underline{z}} \partial_{\underline{z}^{*}}K & = &
{\displaystyle\frac{|\dot{\tau}|^{2}}{(\Im {\rm m}\, \tau)^{2}}}
+{\textstyle\frac{1}{16}} \Im {\rm m}\, \tau\, \mathcal{F}^{2}\, .
  \end{array}
\end{equation}

\item[Case B:] These are well-known solutions determined by a choice of (in
  this case) antiholomorphic function $\tau=\tau(z^{*})$. The vector field
  strengths vanish\footnote{These solutions are given in
    Ref.~\cite{Tod:1995jf} in different coordinates in which the metric
    functions have dependence on $u$, but this dependence can be eliminated.}
  and the metric takes the form

\begin{equation}
ds^{2} = 2 du dv -2e^{2U}dzdz^{*}\, ,
\hspace{1cm}
e^{2U}=\Im {\rm m} (\tau)\, .
\end{equation}

  \end{description}

\end{description}

As for the unbroken supersymmetries of these solutions, they all preserve
generically $1/4$ of the supersymmetries. It is not easy to find generic
conditions for the solutions to preserve $1/2$ (although this has been studied
in special cases, see Ref.~\cite{Bergshoeff:1996gg}). As for maximally
supersymmetric solutions, we only expect Minkowski spacetime, since,
otherwise, there would be another maximally supersymmetric solution of
$N=1,d=10$ supergravity different from 10-dimensional Minkowski spacetime.

The rest of this paper is devoted to proof these results. In
Section~\ref{sec-N4d4} we describe in detail pure, ungauged, $N=4,d=4$
supergravity. In Section~\ref{sec-setup} we define the problem and equations
that we want to solve and find the first consistency conditions. To go on, one
has to consider separately the timelike and null cases. This is done in
Sections~\ref{sec-timelike} and~\ref{sec-null}, respectively.  Our conventions
are described in Appendix~\ref{sec-conventions} and Appendix~\ref{sec-Fierz}
contains all the algebraic identities satisfied by the products of tensors
constructed as bilinears of chiral spinors, derived by Fierzing.


\section{Pure, ungauged, $N=4,d=4$ supergravity}
\label{sec-N4d4}

The bosonic fields of $N=4,d=4$ supergravity multiplet are:

\begin{enumerate}
\item The Einstein metric $g_{\mu\nu}$.
\item The complex scalar $\tau$ that parametrizes an $SL(2,
  \mathbb{R})/U(1)$ coset space. In terms of its real and imaginary
  parts (the axion $a$ and the dilaton $\phi$) it is written
  $\tau=a+ie^{-\phi}$.
\item The 6 $U(1)$ vector fields whose complex combinations we label
  with an antisymmetric pair of $SU(4)$ indices $A_{IJ\, \mu}$,
  $I,J=1,\cdots,4$ and are subject to the reality constraint

  \begin{equation}
  A_{IJ\, \mu}={\textstyle\frac{1}{2}}\varepsilon_{IJKL}A^{KL}{}_{\mu}\, ,
  \end{equation}
  
  where we rise and lower all $SU(4)$ indices by complex conjugation:
  $A^{IJ}{}_{\mu}\equiv (A_{IJ\, \mu})^{*}$. Their field strengths are
  $F_{IJ}= dA_{IJ}$ and are subject to the same reality constraint.

\end{enumerate}

The fermionic fields of this supermultiplet, which are always 4-component
(complex) Weyl spinors, are

\begin{enumerate}
\item The 4 dilatini $\chi_{I}$, which, with lower $SU(4)$ indices, have
  positive chirality.
\item The 4 gravitini $\psi_{I\, \mu}$ which, with lower $SU(4)$ indices, have
  negative chirality.
\end{enumerate}

Complex conjugation raises the $SU(4)$ indices and reverses the chiralities.

There are two global (\textit{hidden}) symmetries in the ungauged theory:
$SU(4)\sim SO(6)$, associated to stringy T~dualities \cite{Maharana:1992my}
and $SL(2,\mathbb{R})$, which is associated to a stringy S~duality
\cite{Font:1990gx}-\cite{Hull:1994ys} and leaves invariant the equations of
motion but not the action.  $SU(4)$ acts on all the fields in the obvious way:

\begin{equation}
\chi^{I\,  \prime}=U^{I}{}_{J}\chi^{J}\, ,
\hspace{1cm}
\chi_{I}{}^{\prime}=\chi_{J}(U^{\dagger})^{J}{}_{I}\, ,
\end{equation}

\noindent
etc. The matrix $\Lambda=\left(
  \begin{array}{cc}
a & b \\
c & d \\
  \end{array}
\right) \in SL(2,\mathbb{R})$ acts on $\tau$ via fractional-linear
transformations

\begin{equation}
\tau^{\prime}=\frac{a\tau +b}{c\tau +d}\, .
\end{equation}

An alternative, linear, description of the action of $\Lambda\in
SL(2,\mathbb{R})$ on $\tau$ can be made using the symmetric $SL(2,\mathbb{R})$
matrix

\begin{equation}
\label{eq:calM1}
\mathcal{M}\equiv \frac{1}{\Im {\rm m}\, \tau}  
\left(
  \begin{array}{rcl}
|\tau|^{2} & \Re {\rm e}\, \tau \\
\Re {\rm e}\, \tau & 1 \\
  \end{array}
\right)\, .
\end{equation}

\noindent
The fractional-linear transformations of $\tau$ are equivalent to the
rule

\begin{equation}
\label{eq:calM2}
\mathcal{M}^{\prime}=\Lambda \mathcal{M} \Lambda^{T}\, .  
\end{equation}

\noindent
Observe that the matrix $S\equiv i\sigma^{2}$ is invariant under
$SL(2,\mathbb{R})$ transformations:

\begin{equation}
\label{eq:Smat}
\Lambda S\Lambda^{T}=S\, . 
\end{equation}

\noindent
The action of $\Lambda\in SL(2,\mathbb{R})$ on the vector fields is
best described by defining the $SL(2,\mathbb{R})$-dual
$\tilde{F}_{IJ}$ of the field strength by

\begin{equation}
\tilde{F}_{IJ}\equiv 
\tau F_{IJ}{}^{+}+ \tau^{*} F_{IJ}{}^{-}= \Re{\rm e} \tau F_{IJ}-\Im{\rm m}
\tau\,  {}^{\star}F_{IJ}\, .
\end{equation}

\noindent
Then, the pair $\tilde{F}_{IJ},\, F_{IJ}$ transforms as an
$SL(2,\mathbb{R})$ doublet, i.e.

\begin{equation}
\label{eq:Fdoublet}
\vec{F}_{IJ}\equiv
\left(
  \begin{array}{c}
\tilde{F}_{IJ}\\ F_{IJ} \\
\end{array}
\right)\, ,
\hspace{1cm}
\vec{F}_{IJ}^{\prime}
=\Lambda \vec{F}_{IJ}\, .  
\end{equation}

\noindent
This implies for $F_{IJ}{}^{\pm}$

\begin{equation}
F^{\prime}_{IJ}{}^{+}= (c\tau +d)F_{IJ}{}^{+}\, ,
\hspace{1cm}
F^{\prime}_{IJ}{}^{-} = (c\tau^{*} +d) F_{IJ}{}^{-}\, .
\end{equation}

Defining the phase of $c\tau +d$ by 

\begin{equation}
e^{2i\varphi}\equiv \frac{c\tau +d}{c\tau^{*} +d}\, ,
\end{equation}

\noindent
we find that, under $SL(2,\mathbb{R})$ several fields and combinations of
fields get a local $U(1)$ phase

\begin{equation}
  \begin{array}{cc}
\chi_{I}^{\prime} = e^{-3i\varphi/2} \chi_{I}\, ,
&
\psi_{I\, \mu}^{\prime} = e^{i\varphi/2} \psi_{I\, \mu}\, ,
\\
& \\
{\displaystyle
\left(\frac{\partial_{\mu} \tau}{\Im {\rm m}\,\tau}\right)^{\prime}
= e^{-2i\varphi} 
\left(\frac{\partial_{\mu} \tau}{\Im {\rm m}\,\tau}\right)\, ,
}
\hspace{.5cm} &
\left[(\Im {\rm m}\,\tau)^{1/2}F_{IJ}{}^{\pm}{}_{\mu\nu}\right]^{\prime}
= e^{\pm i\varphi}
\left[(\Im {\rm m}\,\tau)^{1/2}F_{IJ}{}^{\pm}{}_{\mu\nu}\right]\, , \\
\end{array}
\end{equation}

\noindent
corresponding to $U(1)$ charges $-3,1,-4$ and $\pm 2$ respectively.  The
combination

\begin{equation}
Q_{\mu}\equiv {\textstyle\frac{1}{4}}
\frac{\partial_{\mu} \Re {\rm e}\, \tau}{\Im {\rm m}\,\tau}\, ,
\end{equation}

\noindent
transforms as a $U(1)$ gauge field, $Q_{\mu}^{\prime} =Q_{\mu}
+{\textstyle\frac{1}{2}}\partial_{\mu}\varphi$
and this allows us to define a $U(1)$-covariant derivative 

\begin{equation}
\mathcal{D}_{\mu}=\nabla_{\mu} -iqQ_{\mu}\, ,
\end{equation}

\noindent
acting on fields with $U(1)$ charge $q$. Complex conjugation reverses
chirality and these $U(1)$ charges.

The action for the bosonic fields is

\begin{equation}
  \label{eq:N4d4SUGRAaction}
S=\int d^{4}x\sqrt{|g|}\left[R 
+{\textstyle\frac{1}{2}}
\frac{\partial_{\mu}\tau\,\partial^{\mu}\tau^{*}}{(\Im {\rm m}\,\tau)^{2}}
-{\textstyle\frac{1}{16}} \Im {\rm m}\,\tau F^{IJ\, \mu\nu} F_{IJ\, \mu\nu}
-{\textstyle\frac{1}{16}} \Re {\rm e}\, \tau
F^{IJ\, \mu\nu} {}^{\star}\!\!F_{IJ\, \mu\nu}
\right]\, .  
\end{equation}

\noindent 
It is useful to introduce the following notation for the 
equations of motion of the bosonic fields:

\begin{equation}
\mathcal{E}_{a}{}^{\mu}\equiv 
-\frac{1}{2\sqrt{|g|}}\frac{\delta S}{\delta e^{a}{}_{\mu}}\, ,
\hspace{1cm}
\mathcal{E}\equiv -\frac{2\Im {\rm m}\tau}{\sqrt{|g|}}
\frac{\delta S}{\delta \tau}\, ,
\hspace{1cm}
\mathcal{E}^{IJ\, \mu}\equiv 
\frac{8}{\sqrt{|g|}}\frac{\delta S}{\delta A_{IJ\, \mu}}\, .
\end{equation}

\noindent
Then, the equations of motion take the form

\begin{eqnarray}
\mathcal{E}_{\mu\nu} & = & 
G_{\mu\nu}+{\textstyle\frac{1}{2}}(\Im {\rm m}\,\tau)^{-2}
[\partial_{(\mu}\tau\partial_{\nu)}\tau^{*} 
-{\textstyle\frac{1}{2}}g_{\mu\nu} 
\partial_{\rho}\tau\partial^{\rho}\tau^{*}] 
-{\textstyle\frac{1}{4}}
\Im {\rm m}\,\tau 
F_{IJ}{}^{+}{}_{\mu}{}^{\rho}F^{IJ-}{}_{\nu\rho}\, ,
\label{eq:Emn}\\
& & \nonumber \\
\mathcal{E} & = & \mathcal{D}_{\mu}
\left({\displaystyle\frac{\partial^{\mu}\tau^{*}}{\Im {\rm m}\,\tau}}\right)
-{\textstyle\frac{i}{8}}
\Im {\rm m}\,\tau F^{IJ\,+\, \rho\sigma} F_{IJ}{}^{+}{}_{\rho\sigma}\, ,
\label{eq:E}\\
& & \nonumber \\
\mathcal{E}^{IJ\, \mu} & = & \nabla_{\nu}{}^{\star}\tilde{F}^{IJ\, \nu\mu}\, .
\label{eq:EIJm}
\end{eqnarray}

The Maxwell equation $\mathcal{E}^{IJ\, \mu}$ transforms as an
$SL(2,\mathbb{R})$ doublet together with the Bianchi identity which we denote
for convenience $\mathcal{B}^{IJ\, \mu}$

\begin{equation}
\label{eq:B}
\mathcal{B}^{IJ\, \mu}\equiv \nabla_{\nu}{}^{\star}F^{IJ\, \nu\mu}\, .
\end{equation}

It is easy to see that the combinations 

\begin{equation}
\label{eq:combinations}
\frac{\mathcal{E}_{IJ}{}^{\mu}
-\tau^{*}\mathcal{B}_{IJ}{}^{\mu}}{(\Im {\rm    m}\, \tau)^{1/2}}\, ,
\hspace{1cm}  
\frac{\mathcal{E}_{IJ}{}^{\mu}
-\tau\mathcal{B}_{IJ}{}^{\mu}}{(\Im {\rm    m}\, \tau)^{1/2}}\, ,
\end{equation}

\noindent
have $U(1)$ charges $+2$ and $-2$, respectively. The equation of motion of the
complex scalar $\mathcal{E}$ has $U(1)$ charge $+4$ and the Einstein equation
is neutral.

For vanishing fermions, the supersymmetry transformation rules of the
gravitini and dilatini, generated by 4 spinors $\epsilon_{I}$ of
negative chirality and $U(1)$ charge $+1$, are

\begin{eqnarray}
\label{eq:fermionsusyvariations}
\delta_{\epsilon} \psi_{I\, \mu} & = &
\mathcal{D}_{\mu}\epsilon_{I} 
-{\textstyle\frac{i}{2\sqrt{2}}}  
(\Im {\rm m}\,\tau)^{1/2}F_{IJ}{}^{+}{}_{\mu\nu}\gamma^{\nu}\epsilon^{J}\, ,\\
& & \nonumber \\
\delta_{\epsilon} \chi_{I} & = & 
{\textstyle\frac{1}{2\sqrt{2}}}\frac{\not\!\partial\tau}{\Im {\rm m}\,\tau} 
\epsilon_{I}
-{\textstyle\frac{1}{8}} 
(\Im {\rm m}\,\tau)^{1/2}\!\not\!F_{IJ}{}^{-}\epsilon^{J}\, .
\end{eqnarray}

We also need the supersymmetry transformation rules of the bosonic bosonic
fields, which take the form

\begin{eqnarray}
\delta_{\epsilon} e_{\mu}{}^{a} & = & 
-{\textstyle\frac{i}{4}}( \bar{\epsilon}^{I} \gamma^{a} \psi_{I\, \mu}
+\bar{\epsilon}_{I} \gamma^{a} \psi^{I}{}_{\mu})\, ,
\label{eq:susytranseam}\\ 
& & \nonumber \\
\delta_{\epsilon} \tau  & = & 
-{\textstyle\frac{i}{\sqrt{2}}}
 \Im {\rm m} \tau \bar{\epsilon}^{I} \chi_{I}\, , 
\label{eq:susytranstau}\\ 
& & \nonumber \\
\delta_{\epsilon} A_{IJ\, \mu} & = & 
\frac{\sqrt{2}}{(\Im {\rm m} \tau)^{1/2}} 
\left[
\bar{\epsilon}_{[I} \psi_{J]\, \mu} 
+{\textstyle\frac{i}{\sqrt{2}}} \bar{\epsilon}_{[I}\gamma_{\mu} \chi_{J]} 
+{\textstyle\frac{1}{2}} \epsilon_{IJKL} 
\left(\bar{\epsilon}^{K} \psi^{L}{}_{\mu} + 
{\textstyle{\frac{i}{\sqrt{2}}}} \bar{\epsilon}^{K}
\gamma_{\mu} \chi^{L} \right) 
\right]\, .
\label{eq:susytransaij}
\end{eqnarray}


\section{Supersymmetric configurations: general setup}
\label{sec-setup}

Our goal is to find all the purely bosonic field configurations of $N=4,d=4$
supergravity $\{g_{\mu\nu},A_{IJ\, \mu},\tau, \psi_{I\, \mu}=0,\chi_{I}=0\}$
which are supersymmetric, i.e.~invariant under, at least, one supersymmetry
transformation generated by a supersymmetry parameter $\epsilon_{I}(x)$. Since
the supersymmetry variations of the bosonic fields are odd in fermion fields,
these transformations will always vanish, but the supersymmetry variations of
the fermions, for vanishing fermions, Eqs.~(\ref{eq:fermionsusyvariations}),
may only vanish for special supersymmetry parameters $\epsilon_{I}(x)$
(\textit{Killing spinors}) that solve the \textit{Killing spinor equations}
(KSEs)

\begin{eqnarray}
\delta_{\epsilon} \psi_{I\, \mu} =
\mathcal{D}_{\mu}\epsilon_{I} 
-{\textstyle\frac{i}{2\sqrt{2}}}  
(\Im {\rm m}\,\tau)^{1/2}F_{IJ}{}^{+}{}_{\mu\nu}\gamma^{\nu}\epsilon^{J} & = &
0\, ,
\label{eq:kse1}\\
& & \nonumber \\
2\sqrt{2}\, \delta_{\epsilon} \chi_{I} = 
\frac{\not\!\partial\tau}{\Im {\rm m}\,\tau} 
\epsilon_{I}
-{\textstyle\frac{1}{2\sqrt{2}}} 
(\Im {\rm m}\,\tau)^{1/2}\!\not\!F_{IJ}{}^{-}\epsilon^{J} & = & 0\, .
\label{eq:kse2}
\end{eqnarray}

\noindent
For a known bosonic field configuration these are, respectively differential
and algebraic equations for the Killing spinor, which may or may not exist. We
want to find precisely for which bosonic field configurations these equations
do have at least one solution $\epsilon_{I}$. Our procedure will consist in
assuming the existence of such a solution and derive consistency conditions
for the field configurations.

We shall be talking most of the time about supersymmetric \textit{field
  configurations}. These may or may not be solutions of the classical
equations of motion. There are several conceptual and practical advantages in
doing so. First of all, we would like to emphasize the fact that supersymmetry
does not imply by itself that the equations of motion are solved, although in
general it considerably simplifies the task of solving them. Secondly, it is
sometimes useful to consider that there are external sources for the fields,
out of the regions in which we are solving the equations of motion. Including
those regions with sources implies staying off-shell. Finally, the off-shell
equations of motion of theories with gauge symmetries obey certain gauge
identities. In theories with local supersymmetry and for field configurations
admitting Killing spinors, the gauge identities are known as \textit{Killing
  spinor identities} (KSIs) \cite{Kallosh:1993wx,Bellorin:2005hy} and can be
used either to reduce the number of equations to be explicitly checked or,
having at hands all the off-shell equations of motion of certain field
configuration as we will, they can be used as a consistency check that it is a
supersymmetric field configuration.

Since these identities are the first consistency conditions that can be derived
from the KSEs, we are going to derive them in the next section. We are also
going to see that they are related to the integrability conditions of the KSEs.
then, in Section~\ref{sec-solving} we are going to explain the strategy that
we will follow to find all the supersymmetric configurations.


\subsection{Killing Spinor Identities (KSIs) and integrability conditions of
  the Killing spinor equations}

Using the supersymmetry transformation rules of the bosonic fields
Eqs.~(\ref{eq:susytranseam},\ref{eq:susytranstau}) and (\ref{eq:susytransaij})
we can derive relations between the (off-shell) equations of motion of the
bosonic fields that are satisfied by any field configuration
$\{e^{a}{}_{\mu},A_{IJ\, \mu},\tau\}$ admitting Killing spinors
\cite{Kallosh:1993wx,Bellorin:2005hy}.  These KSIs take, for this theory, the
form

\begin{eqnarray}
i\bar{\epsilon}^{I} \gamma^{a} \mathcal{E}_{a}{}^{\mu} 
+\frac{1}{\sqrt{2}(\Im {\rm m}\tau)^{1/2}}  
\bar{\epsilon}_{J} \mathcal{E}^{\mu JI}  & = & 0\, ,
\label{eq:ksi1}\\ 
& & \nonumber \\
\bar{\epsilon}^{I} \mathcal{E}  
+\frac{1}{\sqrt{2}(\Im {\rm m}\tau)^{1/2}}  
\bar{\epsilon}_{J} \not\! \mathcal{E}^{JI}  & = & 0\, .
\label{eq:ksi2}
\end{eqnarray}

Observe that it is implicitly assumed that the Bianchi identities are
identically satisfied, i.e.

\begin{equation}
\mathcal{B}_{IJ}{}^{\mu}=0\, ,
\end{equation}

\noindent
and, therefore, these identities are not $SL(2,\mathbb{R})$-covariant.  We may
have to take this point into account when comparing with the equations that we
will actually find, but we can also find (with considerably more effort) the
$SL(2,\mathbb{R})$-covariant relations between the equations of motion from
the integrability conditions of the Killing spinor equations (\ref{eq:kse1})
and (\ref{eq:kse2}).

Thus, acting with $\mathcal{D}_{\mu}$ on the Eq.~(\ref{eq:kse1}) using both
Eq.~(\ref{eq:kse1}) and Eq.~(\ref{eq:kse2}) and antisymmetrizing on the vector
indices we get

\begin{equation}
  \begin{array}{rcl}
\mathcal{D}_{[\mu} \delta_{\epsilon} \psi_{I\, \nu]} & = &
{\textstyle\frac{1}{8}}
{\displaystyle\frac{\partial_{[\mu}\tau\partial_{\nu]}\tau^{*}}
{(\Im {\rm m}\, \tau)^{2}}} \epsilon_{I} \\
& & \\
& & 
-{\textstyle\frac{1}{8}}
\left\{R_{\mu\nu}{}^{ab}\delta_{I}{}^{K}
- \Im {\rm m}\, \tau 
F_{IJ}{}^{+}{}_{[\mu}{}^{a}
F^{KJ\, -}{}_{\nu]}{}^{b}
\right\}\gamma_{ab}\epsilon_{K} \\
& & \\
& & 
+{\textstyle\frac{1}{4\sqrt{2}}}(\Im {\rm m}\, \tau)^{-1/2}
\left\{
F_{IJ}{}^{+}{}_{\rho[\nu}\partial_{\mu]}\tau
-2i \Im {\rm m}\, \tau \nabla_{[\mu|}F_{IJ}{}^{+}{}_{|\nu]}
\right\}
\gamma^{\rho}\epsilon^{J} \\
& & \\
& = & 0\, .
\end{array}
\end{equation}

To extract from this integrability condition a relation between the equations
of motion we act with $\gamma^{\nu}$ from the left. We get

\begin{equation}
\label{eq:ks1-2}
4\gamma^{\nu}\mathcal{D}_{[\mu} \delta_{\epsilon} \psi_{I\, \nu]} =
(\mathcal{E}_{\mu\nu} 
-{\textstyle\frac{1}{2}}g_{\mu\nu}\, \mathcal{E}_{\sigma}{}^{\sigma})
\gamma^{\nu}\epsilon_{I} 
-\frac{i}{2\sqrt{2}(\Im {\rm m}\, \tau)^{1/2}}
(\not\! \mathcal{E}_{IJ}  
-\tau^{*}\not\!\!\mathcal{B}_{IJ})
\gamma_{\mu}\epsilon^{J}=0\, .
\end{equation}

Acting now with $\gamma^{\mu}$ and using the result to eliminate
$\mathcal{E}_{\sigma}{}^{\sigma}$ we get, finally
the $SL(2,\mathbb{R})$-covariantization of the KSIs Eq.~(\ref{eq:ksi1})

\begin{equation}
\label{eq:ks2-2}
\mathcal{E}^{\mu}{}_{a}\gamma^{a}\epsilon_{I}    
-\frac{i}{\sqrt{2}(\Im {\rm m}\, \tau)^{1/2}}
(\mathcal{E}_{IJ}{}^{\mu}  
-\tau^{*}\mathcal{B}_{IJ}{}^{\mu})\epsilon^{J}=0\, .
\end{equation}

Similarly, the $SL(2,\mathbb{R})$-covariantization of the KSIs
Eq.~(\ref{eq:ksi1}) can be obtained by calculating
$2\sqrt{2}\not\!\mathcal{D}\delta_{\epsilon}\chi_{I}=0$ and takes the form

\begin{equation}
\label{eq:ks2-3}
\mathcal{E}^{*}\epsilon_{I}  
-\frac{1}{\sqrt{2}(\Im {\rm m}\, \tau)^{1/2}}
(\not\!\!\mathcal{E}_{IJ} -\tau\not\!\!\mathcal{B}_{IJ})\epsilon^{J}=0\, .
\end{equation}

These two identities are now manifestly
$SL(2,\mathbb{R})$-covariant\footnote{See the paragraph after
  Eq.~(\ref{eq:combinations}).}.  The comparison with our results will be
easier if we multiply these equations by gamma matrices and conjugate spinors
$\bar{\epsilon}_{K}$ and $\bar{\epsilon}^{K}$ from the left, to derive
relations involving spinor bilinears. In the case in which the vector $V^{a}$
is timelike, we get

\begin{eqnarray}
\mathcal{E}^{ab} 
-{\textstyle\frac{1}{2}}\Im {\rm m}\, \mathcal{E} V^{a}V^{b} 
-\frac{1}{\sqrt{2}} (\Im {\rm m}\, \tau)^{1/2}
\Im {\rm m}\, (M^{IJ}\mathcal{B}_{IJ}{}^{a})V^{b} & = & 0\, , 
\label{eq:ksi3} \\
& & \nonumber \\
\mathcal{E}^{*} V^{a} -\frac{i}{\sqrt{2}(\Im {\rm m}\, \tau)^{1/2}}
M^{IJ}(\mathcal{E}_{IJ}{}^{a} -\tau\mathcal{B}_{IJ}{}^{a}) 
& = & 0\, ,
\label{eq:ksi4} \\
& & \nonumber \\
\Im {\rm m} [M_{IJ}(\mathcal{E}_{IJ}{}^{a} 
-\tau^{*}\mathcal{B}_{IJ}{}^{a})] & = & 0\, .
\label{eq:ksi5}
\end{eqnarray}

\noindent
Observe that the first equation implies the off-shell vanishing of all the
Einstein equations with one or two spacelike components. Further, the Einstein
equation is automatically satisfied when the Maxwell, Bianchi and complex
scalar equations are satisfied.

When $V^{a}$ is null (we denote it by $l^{a}$), all the spinors $\epsilon_{I}$
are proportional and we can use the parametrization of
Eq.~(\ref{proportional}) in Eqs.~(\ref{eq:ks2-2}) and (\ref{eq:ks2-3}).
Contracting with $\phi^{I}$ using the normalization
Eq.~(\ref{eq:normalization}) and with the conjugate spinors
$\bar{\epsilon},\bar{\epsilon^{*}},\bar{\eta},\bar{\eta^{*}}$, where $\eta$ is
an auxiliary spinor with normalization Eq.~(\ref{eq:auxiliary}), we arrive at
the identities

\begin{eqnarray}
\label{eq:ksi6}
(\mathcal{E}^{\mu}{}_{a}
-{\textstyle\frac{1}{2}}e_{a}{}^{\mu}\mathcal{E}^{\rho}{}_{\rho})\, l^{a} = 
(\mathcal{E}^{\mu}{}_{a}
-{\textstyle\frac{1}{2}}e_{a}{}^{\mu}\mathcal{E}^{\rho}{}_{\rho})\, m^{a} 
& = & 0\, ,\\
& & \nonumber \\
\label{eq:ksi7}
\mathcal{E} & = & 0\, ,\\
& & \nonumber \\
\label{eq:ksi8}
(\mathcal{E}_{IJ}{}^{\mu} 
-\tau^{*}\mathcal{B}_{IJ}{}^{\mu})\phi^{J} & = & 0\, .
\end{eqnarray}

\noindent
where the null complex vectors are defined in Eq.~(\ref{eq:nulltetraddef}).
Observe that in this case supersymmetry implies that the scalar equations of
motion must be automatically satisfied.


\subsection{Solving the Killing spinor equations}
\label{sec-solving}

The procedure we will follow  to find the field configurations for which the
KSEs admit at least one solution will be the following:

\begin{enumerate}
  
\item In Section~\ref{sec-bilinearkse} we are going to reexpress the KSEs as
  differential and algebraic equations for the bilinears (scalars, vectors and
  2-forms, see Appendix~\ref{sec-Fierz}) built with the Killing spinors.
  Solving the equations for all the bilinears is essentially equivalent to
  solving the KSEs.
  
\item In Section~\ref{sec-decompositions} we are going to find, among the
  bilinears, a Killing vector $V^{\mu}$ and decompose the vector field
  strengths w.r.t.~to it computing $V^{\rho}F_{IJ}{}^{+}{}_{\mu\rho}$ or
  $V^{\rho}F_{IJ}{}^{-}{}_{\mu\rho}$ in terms of the scalar bilinears and
  $\tau$ and then using, Eqs.~(\ref{eq:decomposition2}) if $V$ is timelike and
  Eqs.~(\ref{eq:decomposition3}) if $V$ is null. These two cases have to be
  studied separately. One of the reasons is that, in the null case, the field
  strength is not completely determined by its contractions with $V$, but
  there are more differences that we are going to explain shortly and require
  a completely separate analysis.
  
\item In the timelike case, studied in Section~\ref{sec-timelike} we will

  \begin{enumerate}
    
  \item Substitute the expressions of the field strengths in the algebraic
    KSEs ($\delta_{\epsilon}\chi_{I}=0$) to check that it is completely
    solved.
    
  \item Substitute into the equations of motion and we will check whether the
    KSIs Eqs.~(\ref{eq:ksi3},\ref{eq:ksi4},\ref{eq:ksi5}) are indeed
    satisfied or there are additional conditions to be imposed. This is done
    in two steps: first we substitute into the equations of motion of the
    vector fields and the complex scalar which we have already expressed in
    terms of the bilinears in Section~\ref{sec-vectorandscalarequations} and
    then, after we specify the form of the metric in terms of the bilinears,
    we substitute into the Einstein equations in
    Section~\ref{sec-metricequations}. Then we check the KSIs.
    
  \item Substitute, finally, into the differential KSEs
    ($\delta_{\epsilon}\psi_{I\, \mu}=0$) to solve it finding additional
    conditions on the bilinears and the form of the Killing spinors in
    Section~\ref{sec-solvingKSEs}.

\end{enumerate}

The timelike case will be completely solved by then and we will study some
examples.

In the null case, which was completely solved by Tod,

\begin{enumerate}
  
\item As explained in Appendix~\ref{sec-Fierz} all the spinors $\epsilon_{I}$
  are proportional $\epsilon_{I}=\phi_{I}\epsilon$ and we use first this
  information in the KSEs to obtain separate equations for the coefficients
  $\phi_{I}$ and the spinor $\epsilon$. This requires the introduction of a
  $U(1)$ connection $\zeta$ that covariantizes the equations with respect to
  (opposite) local changes of phase of $\phi_{I}$ and $\epsilon$.

\item All the vectors bilinears are also proportional to the Killing vector
  $V^{a}$ which we rename here $l^{a}$. It is convenient to introduce an
  auxiliary spinor to build independent vector bilinears that constitute a
  null tetrad. The KSEs only give partial information about the derivatives of
  these vectors, except for $l^{a}$, which is built with $\epsilon$ and is
  always covariantly constant, the very definition of a $pp$-wave space
  \cite{kn:Br1,kn:Br2}.
  
\item Although the vector field strengths and the derivatives of the vector
  bilinears are not completely determined, it is possible to extract
  information constructing the equations of motion and imposing the KSI. In
  particular we find that the $U(1)$ connection $\zeta$ is trivial.

\item There are two different cases to be considered ($A$ and $B$) which are
  essentially solved by solving first the integrability constraints.

\end{enumerate}

\end{enumerate}


\subsection{Killing equations for the bilinears}
\label{sec-bilinearkse}

We start with the equations $\delta_{\epsilon}\chi_{I}=0$. We just have to
multiply the from the right with gamma matrices and Dirac conjugates of
Killing spinors. We have, in particular, from
$\bar{\epsilon}^{K}\delta_{\epsilon}\chi_{I}=0$

\begin{equation}
\label{eq:vt}
V^{K}{}_{I}\cdot\partial\tau -
{\textstyle\frac{i}{2\sqrt{2}}} 
(\Im {\rm m}\,\tau)^{3/2}F_{IJ}{}^{-}\cdot \Phi^{KJ}=0\, , 
\end{equation}

\noindent
and, from $\bar{\epsilon}^{K}\gamma_{\rho}\delta_{\epsilon}\chi_{I}=0$

\begin{equation}
\label{eq:mt}
F_{IJ}{}^{-}{}_{\rho\sigma}V^{J}{}_{K}{}^{\sigma }
+{\textstyle\frac{i}{\sqrt{2}}} (\Im {\rm m}\,\tau)^{-3/2} 
\left(M_{IK}\partial_{\rho}\tau -
\Phi_{IK\, \rho}{}^{\mu}\partial_{\mu}\tau\right) =0\, . 
\end{equation}

It is possible to derive more Killing equations for the bilinears from
the dilatini supersymmetry rule, but it will not be necessary. 

Let us turn to the gravitini supersymmetry rules. Now we apply
$SL(2,\mathbb{R})$-covariant derivative on the bilinears and use
$\delta_{\epsilon}\psi_{I\, \mu}=0$ to reexpress
$\mathcal{D}_{\mu}\epsilon_{I}$. We get

\begin{eqnarray}
\mathcal{D}_{\mu}M_{IJ} & = & 
{\textstyle\frac{1}{\sqrt{2}}}(\Im {\rm m}\,\tau)^{1/2}  
F_{K[I|}{}^{+}{}_{\mu\nu} V^{K}{}_{|J]}{}^{\nu}\, , \label{eq:dm}\\
& & \nonumber \\
\mathcal{D}_{\mu} V^{I}{}_{J\, \nu} & = & 
-{\textstyle\frac{1}{2\sqrt{2}}}(\Im {\rm m}\,\tau)^{1/2}
\left[M_{KJ}F^{KI\, -}{}_{\mu\nu} +M^{IK}F_{JK}{}^{+}{}_{\mu\nu}
\right. \nonumber\\
& & \nonumber \\
& & \left.
-\Phi_{KJ\, (\mu}{}^{\rho}F^{KI\, -}{}_{\nu)\rho}
-\Phi^{IK}{}_{(\mu|}{}^{\rho}F_{KI}{}^{+}{}_{|\nu)\rho}
\right]\, , \label{eq:dv} \\
& & \nonumber \\
\mathcal{D}_{\mu} \Phi_{IJ\, \mu\nu} & = & 
-{\textstyle\frac{1}{2\sqrt{2}}}(\Im {\rm m}\,\tau)^{1/2}
\left[ 2g_{\mu[\nu|} F_{KI}{}^{+}{}_{|\rho]\alpha} V^{K}{}_{J}{}^{\alpha}
+2F_{KI}{}^{+}{}_{\nu\rho}V^{K}{}_{J\, \mu}
\right. \nonumber \\
& & \nonumber \\
& & 
\left. 
-3 F_{KI}{}^{+}{}_{[\mu\nu|}V^{K}{}_{J\, |\rho]}
+(I\leftrightarrow J)
\right]\, . \label{eq:dp}
\end{eqnarray}


\subsection{First consequences and general results}
\label{sec-decompositions}

Contracting the free indices in Eqs.~(\ref{eq:dv}) and (\ref{eq:vt}) it
is immediate to see that $V^{\mu}\equiv V^{I}{}_{I}{}^{\mu}$ is a
(non-spacelike, Eq.~(\ref{eq:vv3})) Killing vector and

\begin{equation}
\label{eq:vt0}
V^{\mu}\partial_{\mu} \tau =0\, .  
\end{equation}

It is also immediate to prove that 

\begin{equation}
\nabla_{\mu}V^{I}{}_{J}{}^{\mu}=0\, .  
\end{equation}

Let us now consider the implications of the reality constraint of the
vector field strengths on the contraction
$F_{KI}{}^{+}{}_{\mu\nu}V^{K}{}_{J}{}^{\nu}$:

\begin{equation}
F_{KI}{}^{+}{}_{\mu\nu}V^{K}{}_{J}{}^{\nu} = 
{\textstyle\frac{1}{2}}\varepsilon_{KIML}(F_{ML}{}^{-}{}_{\mu\nu})^{*}  
V^{K}{}_{J}{}^{\nu}\, .
\end{equation}

Taking the $SU(4)$ dual in both sides of this equation and taking into
account the reality properties of the vectors $V^{K}{}_{J}{}^{\nu}$,
we get

\begin{equation}
{\textstyle\frac{1}{2}}\varepsilon^{SRIJ}
F_{KI}{}^{+}{}_{\mu\nu}V^{K}{}_{J}{}^{\nu} = 
-{\textstyle\frac{1}{2}}
\left[
F_{SR}{}^{-}{}_{\mu\nu}V^{\nu}
+2 F_{J[S|}{}^{-}{}_{\mu\nu}V^{J}{}_{|R]}{}^{\nu}
\right]^{*}\, ,
\end{equation}

\noindent
from which we get 

\begin{equation}
F_{SR}{}^{-}{}_{\mu\nu}V^{\nu} =
-2 F_{J[S|}{}^{-}{}_{\mu\nu}V^{J}{}_{|R]}{}^{\nu}
-\left[\varepsilon^{SRIJ}
F_{KI}{}^{+}{}_{\mu\nu}V^{K}{}_{J}{}^{\nu}\right]^{*}\, .
\end{equation}

The first and second terms in the r.h.s.~of this equation can be
rewritten in terms of scalars using the antisymmetric part of
Eq.~(\ref{eq:mt}) and the complex conjugate of Eq.~(\ref{eq:dm}). We
get, at last,

\begin{equation}
\label{eq:FSRV}
F_{SR}{}^{-}{}_{\mu\nu}V^{\nu} =
-\frac{\sqrt{2}i}{(\Im {\rm m}\,\tau)^{3/2}}M_{SR}\partial_{\mu}\tau
-\frac{\sqrt{2}}{(\Im {\rm m}\,\tau)^{1/2}}\varepsilon_{SRIJ}
\mathcal{D}_{\mu}M^{IJ}\, .
\end{equation}

\noindent
The complex conjugate of this equation gives us $F^{SR\,
  +}{}_{\mu\nu}V^{\nu}$ and, taking the $SU(4)$-dual we get
$F_{IJ}{}^{+}{}_{\mu\nu}V^{\nu}$ etc. 

From this equation, contracting the free index with $V^{\mu}$ and using
Eq.~(\ref{eq:vt0}) we get immediately

\begin{equation}
\label{eq:vm0}
V^{\mu}\partial_{\mu} M_{IJ} =0\, .  
\end{equation}

Now, the use that we make of this result and the subsequent analysis will
depend on the causal nature if the non-spacelike vector $V^{\mu}$. We must
distinguish between two cases: the case in which it is timelike, which we
consider in section ~\ref{sec-timelike} and the case in which it is null (and
we rename it $l^{\mu}$), which we consider in section~\ref{sec-null}.


\section{The timelike case}
\label{sec-timelike}

If $V^{2}=2M^{IJ}M_{IJ}\equiv 2|M|^{2}\neq 0$ we can use Eq.~(\ref{eq:FSRV})
to express $F_{IJ}{}^{-}$ entirely in terms of scalars, their derivatives, and
$V_{\mu}$ using Eq.~(\ref{eq:decomposition2}):

\begin{equation}
\label{eq:FSR}
F_{SR}{}^{-} =
-\frac{1}{\sqrt{2}|M|^{2}(\Im {\rm m}\,\tau)^{1/2}}
\left\{
\left[i\frac{M_{SR}}{(\Im {\rm m}\,\tau)}d\tau
+\varepsilon_{SRIJ}\mathcal{D}M^{IJ}\right]\wedge \hat{V}
-i\, {}^{\star}\!\left[\cdots\right]\right\}\, .
\end{equation}

Here we have added a hat to $V$ to denote the differential form $\hat{V}\equiv
V_{\mu}dx^{\mu}$ and distinguish its norm.

It can be seen that this form of $F_{SR}{}^{-}$ satisfies identically
all the Killing spinor equations $\delta_{\epsilon}\chi_{I}=0$, that
we can consider solved. 

To solve the equations of motion it is convenient to have directly
$F_{IJ}$ and its $SL(2,\mathbb{R})$-dual $\tilde{F}_{IJ}$. Their
expressions are, actually, somewhat simpler due to the following
property: if $dF=0$ (which is the equation satisfied by $F_{IJ}$ and
$\tilde{F}_{IJ}$) and $\pounds_{V}F=0$ then
$\nabla_{[\mu}(F_{\nu]\rho}V^{\rho})=0$ and, locally,
$F_{\nu\rho}V^{\rho}=\nabla_{\nu}E$ for some scalar potential
$E$. Thus, following  Tod \cite{Tod:1995jf}, we define

\begin{equation}
\nabla_{\mu}E_{IJ}\equiv V^{\nu}F_{IJ\, \nu\mu}\, ,
\hspace{1cm}
\nabla_{\mu}B_{IJ}\equiv 
V^{\nu}\tilde{F}_{IJ\, \nu\mu}\, ,
\end{equation}

\noindent
and, using the above form of $F_{IJ}{}^{-}$ Eq.~(\ref{eq:FSR}) we find

\begin{equation}
\label{eq:potentials}
  \begin{array}{rcl}
E_{IJ} & = & 2\sqrt{2} (\Im {\rm m}\,\tau)^{-1/2} (M_{IJ}+\tilde{M}_{IJ})\, ,\\
& & \\
B_{IJ} & = & 2\sqrt{2} (\Im {\rm m}\,\tau)^{-1/2} 
(\tau M_{IJ} +\tau^{*} \tilde{M}_{IJ}) \, ,\\
\end{array}
\end{equation}

\noindent
where 

\begin{eqnarray}
\label{eq:tildeF}
\tilde{F}_{IJ} & = & 
V^{-2}\left\{ 
\hat{V} \wedge dB_{IJ}
+{}^{\star}\!\!
\left[\hat{V} \wedge\left(\frac{\Re {\rm e}\,\tau}{\Im {\rm m}\,\tau}dB_{IJ} 
-\frac{|\tau|^{2}}{\Im {\rm m}\,\tau} dE_{IJ} \right)\right]
\right\}\, ,\\
& & \nonumber \\
\label{eq:F}
 F_{IJ} & = & 
V^{-2}\left\{ 
 \hat{V}\wedge dE_{IJ}
-{}^{\star}\!\!
\left[\hat{V}\wedge \left(\frac{\Re {\rm e}\,\tau}{\Im {\rm m}\,\tau}dE_{IJ} 
-\frac{1}{\Im {\rm m}\,\tau} dB_{IJ} \right)\right]
\right\}\, .
\end{eqnarray}

It is worth spending a moment in checking the consistency of these
results. By definition, $B_{IJ}$ and $E_{IJ}$ must transform under
$SL(2,\mathbb{R})$ as $\tilde{F}_{IJ}$ and $F_{IJ}$, i.e.~as a
doublet:

\begin{equation}
\vec{E}_{IJ}\equiv
\left(
  \begin{array}{c}
B_{IJ}\\ E_{IJ}\\
\end{array}
\right)\, ,
\hspace{1cm}
\vec{E}_{IJ}^{\prime}
=\Lambda \vec{E}_{IJ}\, .  
\end{equation}

\noindent
We can check that this is consistent with Eqs.~(\ref{eq:tildeF}) and
(\ref{eq:F}) by rewriting the last two equations in the manifestly
$SL(2,\mathbb{R})$-covariant form

\begin{equation}
\vec{F}_{IJ} = V^{-2}\left\{\hat{V}\wedge d\vec{E}_{IJ} 
-{}^{\star}\!\!\left[\hat{V}\wedge (\mathcal{M}Sd\vec{E}_{IJ})\right] 
\right\}\, ,
\end{equation}

\noindent
on account of Eqs.~(\ref{eq:calM1},\ref{eq:calM2}) and
(\ref{eq:Smat}).

On the other hand, it is easy to check that the fact that $\vec{E}_{IJ}$
transforms as a doublet is consistent with the transformations rules of $\tau$
and $M_{IJ}$ alone and Eqs.~(\ref{eq:potentials}).


\subsection{Vector and scalar equations of motion}
\label{sec-vectorandscalarequations}

Our next step consists in finding equations for $M_{IJ}$ and $\tau$ from the
equations of motion using the decompositions of $\tilde{F}_{IJ}$ and $F_{IJ}$
Eqs.~(\ref{eq:tildeF}) and (\ref{eq:F}) in which these fields are written
entirely in terms of those scalars and the Killing vector (1-form) $V$. In
this process we are going to find derivatives of $V$, and we need to express
these in terms of the scalars and $V$ itself.

From Eq.~(\ref{eq:dv}) we find that $V$ satisfies the equation

\begin{equation}
d\hat{V} = -{\textstyle\frac{1}{\sqrt{2}}}(\Im {\rm m}\, \tau)^{1/2}
[M^{IJ}F_{IJ}{}^{+} +M_{IJ}F^{IJ\, -}]\, .  
\end{equation}

\noindent
Since

\begin{equation}
M^{IJ}F_{IJ}{}^{+} = 
-\frac{\sqrt{2}M^{IJ}}{(\Im {\rm m}\, \tau)^{1/2}|M|^{2}}
[\mathcal{D}M_{IJ}\wedge \hat{V} 
+i{}^{\star}\!(\mathcal{D}M_{IJ}\wedge \hat{V})]\, ,
\end{equation}

\noindent
we get 

\begin{equation}
d\hat{V} = \frac{1}{|M|^{2}}\left\{d|M|^{2} \wedge \hat{V}
+i\,{}^{\star}\!\!
\left[(M^{IJ}\mathcal{D}M_{IJ}
-M_{IJ}\mathcal{D}M^{IJ})\wedge \hat{V}\right] \right\}\, .  
\end{equation}

It is also convenient to define the 1-form $\xi$ and the 2-form $\Omega$

\begin{eqnarray}
\xi & \equiv & 
\textstyle{\frac{i}{4}}|M|^{-2}(M_{IJ}dM^{IJ}-M^{IJ}dM_{IJ})\, ,
\label{eq:xidef} \\
& & \nonumber \\
\Omega & \equiv & 
2 |M|^{-2}\, {}^{\star}\!\left[(Q -\xi)\wedge \hat{V}\right]\, .
\end{eqnarray}

\noindent
$\xi$ transforms under $SL(2,\mathbb{R})$ as

\begin{equation} 
\xi' =  \xi +\textstyle{\frac{1}{2}} d \varphi\, ,
\end{equation} 

\noindent
i.e.~as the $U(1)$ connection $Q$, which makes $\Omega$ invariant. The
connection $\xi$ is also orthogonal to $V$ and invariant under local
rescalings of the scalar matrix $M_{IJ}$:

\begin{equation} 
\xi(\Lambda(x) M_{IJ}) = \xi(M_{IJ})\, ,
\end{equation} 

\noindent
a property that we will exploit later on. Further, using Eq.~(\ref{eq:MMJJ})
we can write the curvature of this connection in the form

\begin{equation}
\label{eq:xicurvature}
d\xi = -{\textstyle\frac{i}{2}} d\frac{M^{IJ}}{|M|}\wedge d\frac{M_{KL}}{|M|}
[\delta_{IJ}{}^{KL}
-\mathcal{J}^{K}{}_{[I}\mathcal{J}^{L}{}_{J]}]\, , 
\end{equation}

\noindent 
that relates the triviality of $\xi$ with the constancy of the projection
$\mathcal{J}^{I}{}_{J}$.

Finally, it is convenient to rewrite the equations of motion of the vector and
scalar fields in differential-form language\footnote{We add hats to denote
  differential forms.}:

\begin{eqnarray}
\hat{\vec{\mathcal{E}}}{}^{\, IJ} & \equiv & 
\vec{\mathcal{E}}^{\, IJ}{}_{\mu}dx^{\mu} = 
-{}^{\star}d\vec{F}^{IJ}
=
\left(
  \begin{array}{c}
\hat{\mathcal{E}}{}^{IJ} \\
\hat{\mathcal{B}}{}^{IJ} \\
\end{array}
\right)
\, , \\
& & \nonumber \\
\hat{\mathcal{E}} & \equiv  & \mathcal{E}\hat{V} \, , 
\end{eqnarray}

\noindent
where $\vec{\mathcal{E}}^{IJ}{}_{\mu}$ is the $SL(2,\mathbb{R})$ doublet
formed by the Maxwell and Bianchi identities:

\begin{equation}
\vec{\mathcal{E}}^{IJ\, \mu} \equiv
\left(
  \begin{array}{c}
\mathcal{E}^{IJ\, \mu} \\
\mathcal{B}^{IJ\, \mu}\\
\end{array}
\right)
=
\left(
  \begin{array}{c}
\nabla_{\nu}{}^{\star}\tilde{F}^{IJ\, \nu\mu} \\
\nabla_{\nu}{}^{\star}F^{IJ\, \nu\mu}\\
\end{array}
\right)
\, .  
\end{equation}

Using the expressions that we have found for the Maxwell fields and their
$SL(2,\mathbb{R})$ duals and using the above equation for $dV$ rewritten in
the form

\begin{equation}
\label{eq:dV}
d\hat{V} = \frac{d|M|^{2}}{|M|^{2}} \wedge \hat{V} +2|M|^{2}\Omega\, ,
\end{equation}

\noindent
we find the following two equations for $M_{IJ}$ and $\tau$:

\begin{eqnarray}
{}^{\star}\hat{\vec{\mathcal{E}}}{}^{\, IJ} & = & 
{\textstyle\frac{1}{2}} d\,{}^{\star}\!\!
\left[\frac{\mathcal{M}S d\vec{E}_{IJ}}{|M|^{2}}\wedge \hat{V} \right] 
+d\vec{E}_{IJ}\wedge \Omega\, , \\
& & \nonumber \\
\frac{{}^{\star}\hat{\mathcal{E}}^{*}}{|M|^{2}} & = & 
-\mathcal{D}\, {}^{\star}\!\!
\left[\frac{d\tau}{|M|^{2}\Im {\rm m}\, \tau} \wedge \hat{V}\right] 
+2i \frac{d\tau}{\Im {\rm m}\, \tau}\wedge \Omega 
+2i \frac{\tilde{M}_{IJ}}{|M|^{2}} 
d\, {}^{\star}\! \left(\frac{dM^{IJ}}{|M|^{2}}\wedge \hat{V} \right)\, . 
\label{eq:tequation0}
\end{eqnarray}

These equations can be now be combined (this is the reason behind the
introduction of $V$ into the equation for $\tau$ and the use of differential
forms) and simplified. Using the new variables $N_{IJ}$ defined by

\begin{equation} 
N_{IJ} = \sqrt{\Im {\rm m} \tau} M_{IJ}\, ,
\hspace{1cm}
|N|^{2}= N^{IJ}N_{IJ}=  \Im {\rm m} \tau |M|^{2}\, ,
\end{equation} 

\noindent
we construct a new combination of equations that we call $\hat{a}^{IJ}$

\begin{equation}
\hat{a}^{IJ} \equiv 
\frac{1}{2\sqrt{2}\Im {\rm m} \tau}(\tau \hat{\mathcal{B}}^{IJ}
-\hat{\mathcal{E}}^{IJ})
-{\textstyle\frac{i}{2}}\frac{(N^{IJ}+\tilde{N}^{IJ})}{|N|^{2}}
\hat{\mathcal{E}}^{*}\, , 
\end{equation}

\noindent
and, which, after some massaging, is going to have a much simpler form.  To
present in compact form the equations of motion we define these two equations

\begin{eqnarray}
n^{IJ} & \equiv & (\nabla_{\mu} +4i\xi_{\mu}) 
\left(\frac{\partial^{\mu}N^{IJ}}{|N|^{2}} \right)\, ,
\label{eq:nequation}\\
& & \nonumber \\
e^{*} & \equiv & (\nabla_{\mu} +4i\xi_{\mu}) 
\left(\frac{\partial^{\mu}\tau}{|N|^{2}} \right)\, ,
\label{eq:eequation}
\end{eqnarray}

\noindent
and, in terms of them, we have, switching again from differential form
notation to tensor notation,

\begin{eqnarray} 
a^{IJ} & = & n^{IJ}
-\frac{N^{IJ}+\tilde{N}^{IJ}}{|N|^{2}}\tilde{N}_{KL}n^{KL}\, ,
\label{eq:aequation}\\ 
& & \nonumber \\
\mathcal{B}^{IJ\, a} & = & \sqrt{2}V^{a}
\left\{
\frac{N^{IJ}+\tilde{N}^{IJ}}{|N|^{2}}\Re {\rm e}\, \mathcal{E}
-i(a^{IJ}-\tilde{a}^{IJ})
\right\}\, ,
\label{eq:bianchiequation} \\
& & \nonumber \\
\mathcal{E}^{IJ\, a} & = & \sqrt{2}V^{a}
\left\{
\frac{N^{IJ}+\tilde{N}^{IJ}}{|N|^{2}}\Re {\rm e}\, (\tau\mathcal{E})
-i(\tau^{*}a^{IJ}-\tau\tilde{a}^{IJ})
\right\}\, .
\label{eq:maxwellequation} \\
& & \nonumber \\
\mathcal{E} & = & |M|^{2} e +2i \tilde{N}^{KL}n_{KL}\, . 
\label{eq:tauequation} 
\end{eqnarray}

The combination $|N|^{-2} d \tau $ has $U(1)$ charge $-4$ and, thus, the
second equation is just a $U(1)$-covariant divergence, the covariant
derivative being constructed with the $\xi$ connection.  The first equation
has a similar form and, although $\frac{d N_{IJ}}{|N|^2}$ does not transform
covariantly under $SL(2,\mathbb{R})$, the equation is
$SL(2,\mathbb{R})$-covariant up to terms proportional to the second equation.


\subsection{Metric equations of motion}
\label{sec-metricequations}

These are equations for the scalars $M_{IJ}$ and $\tau$ and involve implicitly
the spacetime metric, which is the only field not determined by them. We need
to study now the Einstein equations and, to do it, it is convenient to choose
coordinates adapted to the timelike Killing vector $V$. We define a time
coordinate by

\begin{equation}
V^{\mu}\partial_{\mu}\equiv\sqrt{2} \partial_{t}\, ,
\end{equation}

\noindent
and the metric takes the ``conformastationary'' form

\begin{equation}
\label{eq:metric}
ds^{2} = |M|^{2}(dt+\omega)^{2} 
-|M|^{-2}\gamma_{\underline{i}\underline{j}}dx^{i}dx^{j}\, ,
\hspace{1cm}
i,j=1,2,3\, ,
\end{equation}

\noindent
where $\omega=\omega_{\underline{i}}dx^{i}$ is a time-independent 1-form and
$\gamma_{\underline{i}\underline{j}}$ is a time-independent
(positive-definite!) metric on constant $t$ hypersurfaces\footnote{The
  components of the connection and curvature of this metric can be found in
  Appendix~\ref{sec-conformastationarymetric}.}. Since $|M|$ is in principle
determined by the above equations, we only need to find equations for $\omega$
and $\gamma$. As usual, the equation for $\omega$ can be derived by comparing
Eq.~(\ref{eq:dV}) for the 1-form $\hat{V}$, with the exterior derivative of
the expression for $\hat{V}$ in the coordinates chosen

\begin{equation}
\hat{V} = \sqrt{2}|M|^{2}(dt +\omega)\, .
\end{equation}

The result is the equation

\begin{equation}
\label{eq:omega}
d\omega = {\textstyle\frac{1}{\sqrt{2}}}\Omega =
 {\textstyle\frac{i}{2\sqrt{2}}}|M|^{-4}\,
 {}^{\star}\!\!\left[(M^{IJ}\mathcal{D}M_{IJ}
 -M_{IJ}\mathcal{D}M^{IJ})\wedge \hat{V}\right]\, .
\end{equation}

Using the conformastationary metric we can reduce all the equations to
equations in the 3 spatial dimensions with the metric $\gamma$. To start with,
the equations $n^{IJ}$ and $e$ defined in Eqs.~(\ref{eq:nequation}) and
(\ref{eq:eequation}) can be expressed in terms of

\begin{eqnarray}
n^{IJ}_{(3)} & \equiv & (\nabla_{\underline{i}} +4i\xi_{\underline{i}}) 
\left(\frac{\partial^{\underline{i}}N^{IJ}}{|N|^{2}} \right)\, ,
\label{eq:nequation3}\\
& & \nonumber \\
e^{*}_{(3)} & \equiv & (\nabla_{\underline{i}} +4i\xi_{\underline{i}}) 
\left(\frac{\partial^{\underline{i}}\tau}{|N|^{2}} \right)\, ,
\label{eq:eequation3}
\end{eqnarray}

\noindent
where all the objects are now 3-dimensional with metric $\gamma$, by 

\begin{equation}
n^{IJ}=-|M|^{2}n^{IJ}_{(3)}\, ,
\hspace{1cm}
e=-|M|^{2}e_{(3)}\, .
\end{equation}

The equation~(\ref{eq:omega}) for the 1-form $\omega$ that enters the
conformastationary metric reduces to

\begin{equation}
\label{eq:omega3}
f_{ij} = 4|M|^{-2}\epsilon_{ijk}
(\xi_{k}-Q_{k})\, ,
\hspace{1cm}
f_{\underline{i}\underline{j}}\equiv
2\partial_{[\underline{i}}\omega_{\underline{j}]}\, .  
\end{equation}

Then, we can express all the equations of motion in terms of these two
equations plus the equation\footnote{This equation should be compared with
  Eq.~(\ref{eq:xicurvature}) in which the antisymmetric part of the same
  combination appears.}

\begin{equation}
e_{ij} \equiv R_{ij}(\gamma) -2\partial_{(i}\left(\frac{N^{IJ}}{|N|}\right)
\partial_{j)}\left(\frac{N_{KL}}{|N|}\right)
(\delta^{KL}{}_{IJ} -\mathcal{J}^{K}{}_{I}\mathcal{J}^{L}{}_{J})\, ,  
\end{equation}

\noindent
as follows:

\begin{eqnarray} 
\mathcal{E}_{00} & = & 
|M|^{2} \left[|M|^{2} \Im {\rm m}e^{*}_{(3)} 
-2\Re {\rm e}\, (N_{KL}n^{KL}_{(3)})
+{\textstyle\frac{1}{2}}e_{k}{}^{k}\right]\, ,
\label{eq:einsteinequation003}\\
& & \nonumber \\
\mathcal{E}_{0i} & = & 0\, ,
\label{eq:einsteinequation0i3}\\
& & \nonumber \\
\mathcal{E}_{ij} & = & |M|^{2}(e_{ij}
-{\textstyle\frac{1}{2}}\delta_{ij}e_{k}{}^{k})\, ,
\label{eq:einsteinequationij3}\\
\mathcal{B}^{IJ\, a} & = & -\sqrt{2}|M|^{2}V^{a}
\left\{
\frac{N^{IJ}+\tilde{N}^{IJ}}{\Im {\rm m}\tau}\Re {\rm e}\, e_{(3)}
-i(n^{IJ}_{(3)}-\tilde{n}^{IJ}_{(3)})
\right\}\, ,
\label{eq:bianchiequation3} \\
& & \nonumber \\
\mathcal{E}^{IJ\, a} & = & -\sqrt{2}|M|^{2}V^{a}
\left\{
\frac{N^{IJ}+\tilde{N}^{IJ}}{\Im {\rm m}\tau}\Re {\rm e}\, 
(\tau e_{(3)})
-i(\tau^{*}n^{IJ}_{(3)}-\tau\tilde{n}^{IJ}_{(3)})
\right\}\, .
\label{eq:maxwellequation3} \\
& & \nonumber \\
\mathcal{E} & = & -|M|^{2} \left[|M|^{2} e_{(3)} 
+2i N_{KL}\tilde{n}^{KL}_{(3)}\right]\, . 
\label{eq:tauequation3} 
\end{eqnarray}

We are now ready to check whether these equations satisfy the relations
expressed in Eqs.~(\ref{eq:ksi3}-\ref{eq:ksi5}). It is immediate to see that 
they do if the following conditions are satisfied off-shell:

\begin{eqnarray}
e_{ij} & = & 0\, ,
\label{eq:susyequation1}\\
& & \nonumber \\
|M|^{2}\Re {\rm e}\,( e_{(3)})
-2\Im {\rm m}(N_{IJ}n^{IJ}_{(3)}) & = & 0\, .
\label{eq:susyequation2}
\end{eqnarray}

The first equation determines the 3-dimensional matric $\gamma$ as a function
of the scalars $N^{IJ}$ and says that $\gamma$ is Ricci-flat is the projection
$\mathcal{J}^{I}{}_{J}$ is constant.  The second equation can be rewritten in
the form

\begin{equation}
\label{eq:susyequation2bis}
\nabla_{\underline{i}}\left(\frac{Q^{\underline{i}}
-\xi^{\underline{i}}}{|M|^{2}}\right)=0\, ,  
\end{equation}

\noindent
and is the integrability condition of Eq.~(\ref{eq:omega3}) for the 1-form
$\omega$, whose existence we have assumed throughout all this analysis. Thus,
it is not so much a necessary condition for supersymmetry as it is a necessary
condition for the whole problem to be well defined.

Let us summarize the results of this section: we have seen that, in the
timelike case at hands, field configurations with a metric of the form
Eq.~(\ref{eq:metric}), vector field strengths of the form Eq.~(\ref{eq:FSR})
and any complex scalar $\tau$, and satisfying Eqs.~(\ref{eq:susyequation1})
and (\ref{eq:susyequation2}) satisfy all the integrability conditions of the
Killing spinor equations.

On the other hand, all the equations of motion, including the Bianchi
identities, are satisfied if the equations

\begin{equation}
e_{(3)}^{*}=0\, ,
\hspace{2cm}
n_{(3)}^{IJ}=0\, ,
\hspace{2cm}
e_{ij}=0\, ,
\end{equation}

\noindent
(were $e_{(3)}^{*}$ and $n_{(3)}^{IJ}$ are defined in
Eq.~(\ref{eq:eequation3}) and Eq.~(\ref{eq:nequation3})) are satisfied, and
automatically the integrability conditions are also satisfied.

We are now ready to check whether the Killing spinor equations always admit
solutions for those field configurations. Thus will help us in solving the
integrability conditions Eqs.~(\ref{eq:susyequation1}) and
(\ref{eq:susyequation2}).


\subsection{Solving the Killing spinor equations}
\label{sec-solvingKSEs}

We have already checked that the equation $\delta_{\epsilon}\chi_{I}=0$ is
automatically solved by our field configurations, and we only have to check
that the equations $\delta_{\epsilon}\psi_{a\, I}=0$ can also be solved for
them.

Let us consider the timelike component first. It can be put in this form:

\begin{equation}
|M|^{-1}
\left\{
\partial_{t}\epsilon_{I} 
-{\textstyle\frac{1}{2}}M^{KL}\mathcal{D}_{i}M_{KL}\gamma_{0i}
\left[\epsilon_{I}+i\sqrt{2}\gamma_{0}\frac{M_{IJ}}{|M|}\epsilon^{J}\right]
+{\textstyle\frac{i}{\sqrt{2}}}|M|
\tilde{\mathcal{J}}^{K}{}_{I}\mathcal{D}_{i}M_{KJ}\gamma_{i}\epsilon^{J}
\right\} =0 \, .
\end{equation}

Using the time-independent projector $\mathcal{J}^{I}{}_{J}$ we can split this equation
into two equations:

\begin{eqnarray}
\partial_{t}\epsilon_{I} 
-{\textstyle\frac{1}{2}}M^{KL}\mathcal{D}_{i}M_{KL}\gamma_{0i}
\left[\epsilon_{I}+i\sqrt{2}\gamma_{0}\frac{M_{IJ}}{|M|}\epsilon^{J}\right] 
& = & 0\, ,\\
& & \nonumber \\
\tilde{\mathcal{J}}^{K}{}_{I}\mathcal{D}_{i}M_{KJ}\gamma_{i}\epsilon^{J}
& = & 0\, . 
\end{eqnarray}

\noindent
The first equation is solved by a time independent spinor because

\begin{equation}
\label{eq:constraint}
\epsilon_{I}+i\sqrt{2}\gamma_{0}\frac{M_{IJ}}{|M|}\epsilon^{J}=0\, ,
\end{equation}

\noindent
due to the Fierz identity

\begin{equation}
M_{IJ}\epsilon^{J}={\textstyle\frac{i}{2}}V^{a}\gamma_{a}\epsilon_{I}\, ,  
\end{equation}

\noindent
and our choice of Vierbeins. For generic (i.e.~not built from already-known
Killing spinors) scalars $M_{IJ}$ the above relation would be a constraint
breaking $1/2$ of the supersymmetries to be imposed on the Killing spinors
whenever $M^{KL}\mathcal{D}_{i}M_{KL}\neq 0$. The counting of unbroken
supersymmetries is, however, a bit more subtle and depends on the triviality
of the $U(1)$ connection $\xi$: if $\xi$ is a total derivative the projection
$\mathcal{J}^{I}{}_{J}$ is constant and a global $SU(4)$ rotation suffices to
set to zero two of the chiral Killing spinors. This is the procedure followed
by Tod in Ref.~\cite{Tod:1995jf}, where he solved the constant
$\mathcal{J}^{I}{}_{J}$ (\textit{internally rigid}) case by setting to zero
two of the spinors, breaking the explicit $SU(4)$ covariance of the solutions.
The solutions found by Tod preserve, then, generically, $1/4$ of the
supersymmetries\footnote{The conditions under which $1/2$ of the
  supersymmetries are preserved were studied in
  Ref.~\cite{Bergshoeff:1996gg}.}. If $\mathcal{J}^{I}{}_{J}$ is not constant,
$\xi$ is non-trivial and the 4 Killing spinors cannot be related by global
$SU(4)$ rotations, but we are now going to see that this case can also be
solved introducing a new projection on the Killing spinors which also reduces
the amount of generically preserved supersymmetries to $1/4$.

Now, using time-independence of the Killing spinors and
Eq.~(\ref{eq:constraint}), the spacelike components of
$\delta_{\epsilon}\psi_{a\, I}=0$ take the form

\begin{equation}
\left[\nabla_{i}
-{\textstyle\frac{1}{2}}\frac{M^{KL}\partial_{i}M_{KL}}{|M|^{2}}\right]
\epsilon_{I}=0\, ,
\end{equation}

\noindent
which can be rewritten in the form

\begin{equation}
\label{eq:xep}
(\nabla_{i} -i\xi_{i})(|M|^{-1/2}\epsilon_{I})=0\, .
\end{equation}

The integrability condition for this equation is 

\begin{equation}
[R_{ijkl}\gamma^{kl} +4i(d\xi)_{ij}]\epsilon_{I}=0\, .
\end{equation}

\noindent
This equation can be solved in essentially one way, up to local Lorentz
transformations:

\begin{equation}
R_{12}{}^{12}=\pm 2 (d\xi)_{12}\, ,
\hspace{2cm}
{\textstyle\frac{1}{\sqrt{2}}}(1\mp i\gamma_{12})\epsilon_{I}=0\, ,
\end{equation}

\noindent
the remaining components of the curvatures being zero. In terms of the
connections we should have, in the appropriate Lorentz frame, the following
relation between the 3-dimensional spin connection $o^{ij}$ and the $U(1)$
connection $\xi$:

\begin{equation}
\xi = \pm {\textstyle\frac{1}{2}} o^{12} (x^{1},x^{2}) 
+ {\textstyle\frac{1}{2}}d\lambda  (x^{1},x^{2},x^{3})\, ,
\end{equation}

\noindent
for some 3-dimensional 1-form $\zeta$ and some real scalar function $\lambda$.
If complex scalars $M^{IJ}$ and 3-dimensional metric
$\gamma_{\underline{i}\underline{j}}$ exist such that the above condition is
met, then there are Killing spinors of the form

\begin{equation}
\epsilon_{I}= e^{\frac{i}{2}\lambda} |M|^{1/2}\epsilon_{I}\, ,
\hspace{1cm}
(\xi - {\textstyle\frac{1}{2}}d\lambda){\textstyle\frac{1}{\sqrt{2}}}(1\mp
i\gamma_{12})\epsilon_{I}=0\, .
\end{equation}

The relation between the spin connection and the $U(1)$ connection is just the
requirement that the 3-dimensional metric has $U(1)$ holonomy, which implies
that it is reducible to the direct product of a 2- and a 1-dimensional metric
and, thus, can always be written in the form

\begin{equation}
\gamma_{\underline{i}\underline{j}}dx^{\underline{i}}dx^{\underline{j}}
= dx^{2}+2e^{2U(z,z^{*})}dzdz^{*}\, ,
\end{equation}

\noindent
which, in turn, implies that $\xi$ is given by

\begin{equation}
\label{eq:xiU}
\xi = \pm{\textstyle\frac{i}{2}}
(\partial_{\underline{z}}Udz -\partial_{\underline{z}^{*}}Udz^{*}) 
+{\textstyle\frac{1}{2}}d\lambda  (x,z,z^{*})\, .  
\end{equation}

Let us summarize the results of this section.  We have found that, to
construct a supersymmetric configuration (not necessarily a solution) of pure,
ungauged, $N=4,d=4$ supergravity amounts, now, to

\begin{enumerate}
\item Find a set of time-independent complex scalars $M^{IJ}$ satisfying
  $\varepsilon^{IJKL}M_{IJ}M_{KL}=0$ such that the $U(1)$ connection $\xi$
  defined in Eq.~(\ref{eq:xidef}) can be written in the form
  Eq.~(\ref{eq:xiU}). The integrability condition Eq.~(\ref{eq:susyequation1})
  should automatically be solved by this choice.
  
\item Find $\tau$ by solving the integrability condition
  Eq.~(\ref{eq:susyequation2bis}) of the defining equation of the 1-form
  $\omega$ (\ref{eq:omega}).

\end{enumerate}

If we want the supersymmetric configuration to be a solution of the equations
of motion, we also need to impose Eqs.~(\ref{eq:nequation3}) and
(\ref{eq:eequation3}), but we do not need to check the integrability condition
Eq.~(\ref{eq:susyequation2bis}).

In the next section we study different solutions to these equations.


\subsection{Supersymmetric configurations and solutions}
\label{sec-susyconfigsolu}

According to the recipe of the previous section, our first step in finding
supersymmetric configurations and solutions is to find the complex scalars
$M^{IJ}$ satisfying $\varepsilon^{IJKL}M_{IJ}M_{KL}=0$ and such that $\xi$ can
be written in the form Eq.~(\ref{eq:xiU}). The first condition can be easily
met, for instance, by taking only $M_{12},M_{13}$ and $M_{23}$ non-vanishing,
but we prefer not to make any specific choice that would break $SU(4($
covariance.  The second condition can be solved by the following
\textit{Ansatz}

\begin{equation}
\label{eq:ansatzm}
M_{IJ}= e^{i\lambda (x,z,z^{*})} M (x,z,z^{*}) k_{IJ}(z)\, ,  
\hspace{.5cm}
M=M^{*}\, ,
\hspace{.5cm}
\lambda=\lambda^{*}\, ,
\hspace{.5cm}
\varepsilon^{IJKL}k_{IJ}k_{KL}=0\, ,
\end{equation}

\noindent
which give a connection $\xi$ of the form  Eq.~(\ref{eq:xiU}) with 

\begin{equation}
U= +\ln{|k|}\, ,
\hspace{1cm}
|k|^{2}\equiv k^{IJ}(z^{*})k_{IJ}(z)\, , 
\end{equation}

\noindent
and satisfies automatically the integrability condition
Eq.~(\ref{eq:susyequation1}). 

Solving the integrability condition Eq.~(\ref{eq:susyequation2bis}) is
considerably more difficult and considering solutions (instead of general
configurations) simplifies the problem. We have found three families of
solutions.

\begin{enumerate}
  
\item If the $k_{IJ}$ are constants, then, normalizing $|k|^{2}=1$ for
  simplicity, $\xi=\frac{1}{2}d\lambda$ and $U=0$.  This is the case
  considered by Tod in Ref.~\cite{Tod:1995jf} and studied in detail in
  Ref.~\cite{Bergshoeff:1996gg}. Tod took advantage of the fact that $d\xi=0$
  implies that $\mathcal{J}^{I}{}_{J}$ is constant and a global $SU(4)$
  rotation can be used to set to zero two of the $\epsilon_{I}$s. We will not
  do so, as this breaks the explicit $SU(4)$ covariance, but our results are,
  of course, equivalent.
  
  Eq.~(\ref{eq:nequation3}) takes the form

  \begin{equation}
\partial_{\underline{i}}  \partial_{\underline{i}} \mathcal{H}_{1}=0\, ,
\hspace{1cm}
 \mathcal{H}_{1}\equiv
[(\Im {\rm m}\, \tau)^{1/2} e^{-i\lambda}M]^{-1}\, ,
  \end{equation}

\noindent
and is solved by any arbitrary complex harmonic function $\mathcal{H}_{1}$.

Using the above equation, Eq.~(\ref{eq:eequation3}) takes the form 

\begin{equation}
\partial_{\underline{i}}\partial_{\underline{i}}( \mathcal{H}_{1}\tau)=0\, ,  
\end{equation}

\noindent
which is solved by 

\begin{equation}
\tau= \mathcal{H}_{1}/ \mathcal{H}_{2}\, ,
\hspace{1cm} 
\partial_{\underline{i}} \mathcal{H}_{2}=0\, , 
\end{equation}

\noindent
another arbitrary complex harmonic function. The pair of harmonic functions
and the constants determine completely the solutions. In particular

\begin{equation}
\label{eq:Meq}
|M|^{-2}=M^{-2}=\Im {\rm m}(\bar{\mathcal{H}_{2}}\mathcal{H}_{1})\, .  
\end{equation}

\item If $e^{i\lambda}=M=1$, the integrability condition
  Eq.~(\ref{eq:susyequation2bis}) can be solved by taking $\tau$ constant.
  The only non-trivial equation of motion, Eq.~(\ref{eq:nequation3}) is solved
  using the holomorphicity of the $k_{IJ}$s. The metric takes the form

\begin{equation}
ds^{2}=|k|^{2}(dt +\omega_{\underline{x}}dx) -|k|^{-2}dx^{2} -2dzdz^{*}\, ,    
\end{equation}

\noindent
where $\omega_{\underline{x}}$ satisfies

\begin{equation}
\partial_{\underline{z}}\omega_{\underline{x}}
-\partial_{\underline{x}}\omega_{\underline{z}} = 
\partial_{\underline{z}^{*}}|k|^{-2}\, ,
\hspace{.5cm}  
\partial_{\underline{z}^{*}}\omega_{\underline{x}}
-\partial_{\underline{x}}\omega_{\underline{z}^{*}} = 
\partial_{\underline{z}}|k|^{-2}\, ,
\hspace{.5cm}  
\partial_{\underline{z}^{*}}\omega_{\underline{z}}
-\partial_{\underline{z}}\omega_{\underline{z}^{*}}=0\, .
\end{equation}

The metric and the supersymmetry projectors indicate that these solutions
describe stationary strings lying along the coordinate $x$, in spite of the
trivial axion field, which is the dual of the Kalb-Ramond 2-form $B$ that
couples to strings.  Observe, however, that the duality relation is not simply
$dB={}^{\star}da$: there are terms quadratic in the field strengths involved
in the duality which must render $B$ non-trivial. 

The metric the the vector fields involved depends strongly on the choice of
holomorphic $k_{IJ}$s. It is instructive to have an example completely worked
out.

Let us consider the simplest case: only $k_{12}=\frac{1}{\sqrt{2}z}$
non-trivial. This allows us to set
$\omega_{\underline{z}}=\omega_{\underline{z}^{*}}=0$.  Then,
$|k|^{2}=|z|^{-2}$ and $\omega_{\underline{x}}=2\Re{\rm e}(z^{2})$ and the
full solution is given by

\begin{equation}
  \begin{array}{rcl}
ds^{2} & = & {\displaystyle\frac{1}{|z|^{2}}}
[dt+2\Re{\rm e}(z^{2})dx]^{2} -|z|^{2}dx^{2} -2dzdz^{*}\, , \\
& & \\
F_{12} & = & -{\displaystyle\frac{\sqrt{2}e^{\phi_{0}/2}}{z^{2}}}
\{[dt+2\Re{\rm e}(z^{2})dx]\wedge dz 
-i{}^{\star}[[dt+2\Re{\rm e}(z^{2})dx]\wedge dz]\}
=(F_{34})^{*}\, ,\\
& & \\
\tau & = & \tau_{0}\, .\\   
\end{array}
\end{equation}

\item The only solutions that we have found with $\lambda$ and the
  $k_{IJ}(z)$s simultaneously nontrivial have just $\lambda=\lambda(x)$ and
  $M=M(x)$ and are a superposition of the solutions with constant $k_{IJ}$ and
  the solutions with constant $\lambda$ in which these functions depend only
  on mutually transversal directions. 
  
  Thus, these solutions depend on holomorphic functions $k_{IJ}(z)$ chosen
  with the same criteria as in the previous case, and a pair of complex
  functions $\mathcal{H}_{1},\mathcal{H}_{2}$ linear in $x$ such that $\Im
  {\rm m}\, \tau > 0$, and the metric is given by

\begin{equation}
ds^{2}=(M|k|)^{2}(dt +\omega_{\underline{x}}dx) 
-(M|k|)^{-2}dx^{2} -2M^{-2}dzdz^{*}\, ,    
\end{equation}

\noindent
where $M$ is again given by Eq.~(\ref{eq:Meq}).

\end{enumerate}


\section{The null case}
\label{sec-null}

As we have mentioned before, the null case was completely solved by Tod in
Ref.~\cite{Tod:1995jf}, but we include it her for the sake of completeness.

As explained in Appendix~\ref{sec-Fierz}, in the null case all the spinors a
proportional $\epsilon_{I}=\phi_{I}\epsilon$. In the $N=4,d=4$ case at hands,
$\epsilon_{I}$ has a $U(1)$ charge under $SL(2,\mathbb{R})$ transformations
that has to be distributed between $\phi_{I}$ and $\epsilon$. We choose to
have the $\phi_{I}$ uncharged. Had we chosen to have $\phi_{I}$ is charged
with charge $q_{\phi}\neq 0$, then the real 1-form

\begin{equation}
\zeta\equiv i\phi_{I} d\phi^{I}\, ,
\end{equation}

\noindent
would transform as a $U(1)$ connection under $SL(2,\mathbb{R})$
transformations as well and would play a role analogous to that of the
connection $\xi$ in the timelike case. With our choice, $\zeta$ is just a
$U(1)$ connection under the transformations Eq.~(\ref{freephase}) and
covariantizes with respect to them the expressions that involve $\epsilon$.

We are now going to substitute $\epsilon_{I}=\phi_{I}\epsilon$ into the KSEs
and we are going to use the normalization condition to split the KSEs into
three algebraic and one differential equation for $\epsilon$. One of the
algebraic equations for $\epsilon$ will be a differential equation for
$\phi_{I}$.

The substitution yields immediately

\begin{eqnarray}
\mathcal{D}_{\mu} \phi_{I} \epsilon+\phi_{I} \mathcal{D}_{\mu}\epsilon 
-{\textstyle\frac{i}{2\sqrt{2}}} 
(\Im{\rm m}\, \tau)^{1/2} F_{IJ}{}^{+}{}_{\mu\nu}\phi^{J}
\gamma^{\nu} \epsilon^{*} & = & 0\, , 
\label{eq:gravitinodegenerate} \\
& & \nonumber \\
\phi_{I}\frac{\not\!\partial\tau}{\Im{\rm m}\, \tau} \epsilon 
-{\textstyle\frac{1}{2\sqrt{2}}} (\Im{\rm m}\, \tau)^{1/2}
\not\!\!  F_{IJ}{}^{-} \phi^{J}\epsilon^{*} & = & 0 \, . 
\label{eq:dilatinodegenerate}
\end{eqnarray}

Acting on Eq.~(\ref{eq:gravitinodegenerate}) with $\phi^{I}$ leads to

\begin{equation}
\mathcal{D}_{\mu}\epsilon = -\phi^{I} \mathcal{D}_{\mu} \phi_{I} \epsilon\, ,
\end{equation}

\noindent
which takes the form

\begin{equation}
\label{eq:Depsilon}
\tilde{\cal D}_{\mu}\epsilon\equiv 
(\mathcal{D}_{\mu} +i\zeta_{\mu})\epsilon=0\, ,  
\end{equation}

\noindent
and becomes the only differential equation for $\epsilon$.  We have defined
the derivative $\tilde{\cal D}$ covariant with respect to $SL(2,\mathbb{R})$
and $U(1)$ local rotations under which $\epsilon$ and $\phi_{I}$ have charges
$+1$ and $-1$, respectively. Using Eq.~(\ref{eq:Depsilon}) into
Eq.~(\ref{eq:gravitinodegenerate}) to eliminate $\mathcal{D}_{\mu}\epsilon$ we
obtain

\begin{equation}
\label{eq:gravitinodegII}
\tilde{\cal D}\phi_{I}
\epsilon  - {\textstyle\frac{i}{2\sqrt{2}}} 
(\Im{\rm m}\, \tau)^{1/2} F_{IJ}{}^{+}{}_{\mu\nu}\phi^{J}\gamma^{\nu} 
\epsilon^{*}= 0\, ,
\end{equation}

\noindent
which is one of the algebraic constraints for $\epsilon$ and is a differential
equation for $\phi_{I}$.

Acting with $\phi^{I}$ on Eq.~(\ref{eq:dilatinodegenerate}) we see that it
splits into two algebraic constraints for $\epsilon$:

\begin{eqnarray}
\not\!\partial\tau \epsilon & = & 0\, , \label{eq:dtaueps} \\
& & \nonumber \\
\not\!\!F_{IJ}{}^{-} \phi^{J}\epsilon^{*} & = & 0\, . \label{eq:vinF-}
\end{eqnarray}

Finally, we add to the system an auxiliary spinor $\eta$, introduced in
Appendix~\ref{sec-Fierz}, with charges opposite to those of $\epsilon$.  The
normalization condition Eq.~(\ref{eq:normalization}) will be preserved if and
only if $\eta$ satisfies a differential equation of the form

\begin{equation}
\label{eq:Deta}
\tilde{D}_{\mu}\eta +a_{\mu}\epsilon=0\, ,  
\end{equation}

\noindent
where $a_{\mu}$ is, in principle, an arbitrary vector with the right
charges that transforms under the redefinitions Eqs.~(\ref{eq:redef})
and (\ref{eq:redef2}) as a connection

\begin{equation}
a_{\mu}^{\prime}=a_{\mu}+\partial_{\mu}\delta\, .
\end{equation}

In practice, however, $a_{\mu}$ cannot be completely arbitrary since the
integrability conditions of the differential equation of $\eta$ have to be
compatible with those of the differential equation for $\epsilon$ and this
requirement will determine $a_{\mu}$.

Before we start a systematic analysis of these equations, it is worth
comparing Eq.~(\ref{eq:Depsilon}) to Eq.~(\ref{eq:xep}) and their
integrability conditions which have the same structure except for the
important detail of the dimensionality and signature. Therefore, we expect two
main types of solutions: configurations with $U(1)$ holonomy on a
2-dimensional (spacelike) subspace and configurations with $U(1)$ holonomy in
a null direction, which is the new possibility allowed by the Lorentzian
signature. These expectations are also supported by the Fierz identities

\begin{eqnarray}
\not\! m \epsilon  & = & -i\epsilon\, ,\\
& & \nonumber \\
\not l \epsilon^{*} & = & 0\, ,
\end{eqnarray}

\noindent
which are satisfied automatically here, but will be interpreted as
projections. 

We will call these two possibilities $B$ and $A$ respectively.


\subsection{Killing equations for the vector bilinears and first consequences}

We are now ready to derive equations involving the bilinears, in particular
the vector bilinears which we construct with $\epsilon$ and the auxiliary
spinor $\eta$ introduced in Appendix~\ref{sec-Fierz}. First we deal with the
equations that do not involve derivative of the spinors. Acting with
$\bar{\epsilon}$ on Eq.~(\ref{eq:gravitinodegII}) and with
$\bar{\epsilon^{*}}\gamma^{\mu}$ on the complex conjugate of
Eq.~(\ref{eq:vinF-}) we get

\begin{eqnarray}
\label{eq:lF1}
  \phi^{I} F_{IJ}{}^{+}{}_{\mu\nu}l^{\nu} & = & 0\, ,\\
& & \nonumber \\
\label{eq:lF2}
\epsilon^{IJKL}\phi_{J} F_{KL}{}^{+}{}_{\mu\nu}l^{\nu} & = & 0\, .
\end{eqnarray}

\noindent
Acting with $\bar{\epsilon^{*}}$ and $\bar{\eta^{*}}$ on
Eq.~(\ref{eq:dtaueps}) we get\footnote{The first of these equations had
  already been obtained in the general case Eq.~(\ref{eq:vt0}).}

\begin{eqnarray}
l\cdot\partial\tau & = & 0\, , \label{eq:ldt}\\
& & \nonumber \\
m^{*}\cdot\partial\tau & = & 0\, . \label{ldm}
\end{eqnarray}

Now, from Eqs.~(\ref{eq:Depsilon}) and (\ref{eq:Deta}) we find

\begin{eqnarray}
\label{eq:dtetrad1}
  \nabla_{\mu} l_{\nu} & = & 0\, ,\\
& & \nonumber \\
\label{eq:dtetrad2}
  \tilde{\cal D}_{\mu} n_{\nu} & = & 
-a^{*}_{\mu}m_{\nu} -a_{\mu}m^{*}_{\nu}\, ,\\
& & \nonumber \\
\label{eq:dtetrad3}
  \tilde{\cal D}_{\mu} m_{\nu} & = & -a_{\mu}l_{\nu}\, .
\end{eqnarray}

Let us now find the simplest implications of these equations.

To start with,Eqs.~(\ref{eq:lF1}) and (\ref{eq:lF2}), together, imply for
nonvanishing $\phi_{I}$\footnote{This equation also follows from the general
  result Eq.~(\ref{eq:FSR}) for vanishing scalars $M_{IJ}$.}

\begin{equation}
F_{IJ}{}^{+}{}_{\mu\nu}l^{\nu} = 0\, .
\end{equation}

Using Eq.~(\ref{eq:decomposition3}), we see that the vector field strengths
must take the form

\begin{eqnarray}
F_{IJ}{}^{+} & = & {\textstyle\frac{1}{2}}
\mathcal{F}_{IJ}\,  l\wedge m^{*}\, , \label{eq:F+deg}\\
& & \nonumber \\
F_{IJ}{}^{-} & = &  {\textstyle\frac{1}{2}}
\tilde{\cal F}_{IJ}\,  l\wedge m\, ,
\end{eqnarray}

\noindent
where $\mathcal{F}_{IJ}$ is a skew-symmetric $SU(4)$ matrix of scalars to be
determined and $\tilde{\cal F}_{IJ}$ is its $SU(4)$ dual.

This solves completely Eq.~(\ref{eq:vinF-}), as can be seen using the Fierz
identity

\begin{equation}
l_{\mu}\gamma^{\mu\nu}\epsilon^{*}=3l^{\nu}\epsilon^{*}\, ,
\end{equation}

\noindent
and we can substitute Eq.~(\ref{eq:F+deg}) into Eq.~(\ref{eq:gravitinodegII})
the only remaining equation in which vector field strengths occur. Using the
Fierz identities

\begin{eqnarray}
\not l \epsilon^{*} & = & 0\, ,\\
& & \nonumber \\
\not\! m^{*} \epsilon^{*} & = & -i\epsilon\, ,
\end{eqnarray}

\noindent
it takes the form

\begin{equation}
\label{eq:residuogravitino}
\tilde{\cal D}_{\mu}\phi_{I}
-{\textstyle\frac{1}{4\sqrt{2}}}
(\Im{\rm m}\, \tau)^{1/2}\mathcal{F}_{IJ}\phi^{J} l_{\mu}
=0\, ,
\end{equation}

\noindent
from which we find

\begin{equation}
  \label{eq:FIJ}
\mathcal{F}_{IJ}\phi^{J}
= \frac{4\sqrt{2}}{(\Im{\rm m}\, \tau)^{1/2}}
n^{\mu}\tilde{\cal D}_{\mu}\phi_{I}\, .
\end{equation}

On the other hand, from Eqs.~(\ref{eq:ldt}) and (\ref{ldm}) we find that

\begin{equation}
d\tau = A\hat{l} +B\hat{m}^{*}\, .  
\end{equation}

There are two cases to be considered here: case $A$ ($B=0$) and case $B$
($B\neq 0$). In case $B$, we can write

\begin{equation}
d\tau = B\left(\hat{m}^{*}+\frac{A}{B}\, \hat{l}\right)
=B \hat{m}^{* \, \prime} \, ,  
\end{equation}

\noindent
after a redefinition of the type Eqs.~(\ref{eq:redef}) and (\ref{eq:redef2}).
All the equations that we have written so far are covariant with respect to
this kind of transformations and we just have to add primes (which we suppress
immediately afterwards) everywhere. Thus, the case $B$ is equivalent to $A=0$
and we can always assume that either $A$ or $B$ is always zero.  Since the
connection $Q$ depends on $\tau$, the holonomy is different in these two
cases. These are the two cases we mentioned at the end of the previous
section and we will deal with them separately afterwards.


\subsection{Equations of motion and integrability constraints}

Although we have not yet discussed the form of the metric, we already have
enough information to study the equations of motion and check whether they
satisfy the integrability conditions Eqs.~(\ref{eq:ksi6})-(\ref{eq:ksi8}).

Using the results of the previous section, we can write the equations of
motion in the form\footnote{We have ignored all the terms that contain
  products $AB$ etc.}

\begin{eqnarray}
\mathcal{E}_{\mu\nu}
-{\textstyle\frac{1}{2}}g_{\mu\nu}\mathcal{E}^{\rho}{}_{\rho}
& = & 
R_{\mu\nu} 
+\left[
{\displaystyle\frac{|A|^{2}}{2(\Im {\rm m}\, \tau)^{2}}}
+{\textstyle\frac{1}{16}}\Im {\rm m}\, \tau\, \mathcal{F}^{2}
\right]l_{\mu}l_{\nu}
+{\displaystyle\frac{|B|^{2}}{2(\Im {\rm m}\,
    \tau)^{2}}}m_{(\mu}m^{*}_{\nu)}\, , \\
& & \nonumber \\
\mathcal{E} 
& = &
{\displaystyle\frac{1}{\Im {\rm m}\, \tau}}
\left[
l^{\mu}\partial_{\mu}A^{*} 
-B^{*} l^{\mu}a_{\mu}
+m^{\mu}\partial_{\mu}B^{*} 
+{\textstyle\frac{i}{4}}
{\displaystyle\frac{|B|^{2}}{\Im {\rm m}\, \tau}}
\right]\, ,\\
& & \nonumber \\
\hat{\mathcal{E}}_{IJ} 
-\tau^{*}\hat{\mathcal{B}}_{IJ} 
& = & 
-i(\Im {\rm m}\, \tau)\, d(\mathcal{F}_{IJ} 
\hat{l}\wedge \hat{m}^{*})\, .
\end{eqnarray}

Substituting into Eqs.~(\ref{eq:ksi6})-(\ref{eq:ksi8}) and operating, we get

\begin{eqnarray}
\label{eq:ksi9}
R_{\mu\nu} l^{\nu} & = & 0\, ,\\
& & \nonumber \\
\label{eq:ksi10}
R_{\mu\nu} m^{\nu} 
-{\displaystyle\frac{|B|^{2}}{4(\Im {\rm m}\, \tau)^{2}}}m_{\mu}
& = & 0\, ,\\
& & \nonumber \\
l^{\mu}\partial_{\mu}A^{*} 
-B^{*} l^{\mu}a_{\mu}
+m^{\mu}\tilde{\mathcal{D}}_{\mu}B^{*} 
+{\textstyle\frac{i}{4}}
{\displaystyle\frac{|B|^{2}}{\Im {\rm m}\, \tau}}
& = & 0\, ,\\
& & \nonumber \\
\label{eq:ksi11}
B^{*}\mathcal{F}_{IJ}\phi^{J} & = & 0\, .
\end{eqnarray}

We do not have a metric yet, but we can find $R_{\mu\nu} l^{\nu}$ and
$R_{\mu\nu} m^{\nu}$ from the integrability conditions of
Eqs.~(\ref{eq:Depsilon}) and (\ref{eq:Deta}). Commuting the derivative and
projecting with gamma matrices and spinors in the usual way, it is easy to
find from Eq.~(\ref{eq:Depsilon})

\begin{eqnarray}
R_{\mu\nu} l^{\nu} & = & -2i (d\zeta)_{\mu\nu}l^{\nu}\, ,\\
& & \nonumber \\
R_{\mu\nu} m^{\nu} & = & +2i (d\zeta)_{\mu\nu}m^{\nu}-2i
(dQ)_{\mu\nu}m^{\nu}\nonumber \\
& & \nonumber \\
& = & +2i (d\zeta)_{\mu\nu}m^{\nu}
+{\displaystyle\frac{|B|^{2}}{4(\Im {\rm m}\, \tau)^{2}}}m_{\mu}
\, ,
\end{eqnarray}

\noindent
and from Eq.~(\ref{eq:Deta})

\begin{eqnarray}
R_{\mu\nu} m^{\nu} & = & 2i (d\zeta)_{\mu\nu}m^{\nu}
-2i (dQ)_{\mu\nu}m^{\nu}+2(da)_{\mu\nu}l^{\nu}
\nonumber \\
& & \nonumber \\
& = & +2i (d\zeta)_{\mu\nu}m^{\nu}
+{\displaystyle\frac{|B|^{2}}{4(\Im {\rm m}\, \tau)^{2}}}m_{\mu}
+2(da)_{\mu\nu}l^{\nu}\, ,\\
& & \nonumber \\
R_{\mu\nu} n^{\nu} & = & 2i (d\zeta)_{\mu\nu}n^{\nu}
-2i (dQ)_{\mu\nu}n^{\nu}+2(da)_{\mu\nu}m^{*\, \nu}\nonumber \\
& & \nonumber \\
& = & 
2i (d\zeta)_{\mu\nu}n^{\nu} + 2(da)_{\mu\nu}m^{*\, \nu}\, .
\end{eqnarray}

Comparing now these three sets of equations, we get 

\begin{equation}
(d\zeta)_{\mu\nu}l^{\nu} = (d\zeta)_{\mu\nu}m^{\nu} =0\, ,\,\,\,\,
\Rightarrow
d\zeta=0\, ,\,\,\,\,
\Rightarrow \zeta = d\alpha\, ,
\end{equation}

\noindent
locally, and, eliminating $\zeta$ by a local phase redefinition,

\begin{eqnarray}
(da)_{\mu\nu} l^{\nu} & = &  0\, ,\\
& & \nonumber \\
(da)_{\mu\nu} m^{*\, \nu} & = &  
-{\textstyle\frac{1}{2}}R_{\mu\nu}n^{\nu}\, ,
\end{eqnarray}

\noindent
which tell us that 

\begin{equation}
\label{eq:da}
da= -{\textstyle\frac{1}{2}} R_{z^{*}u}
\hat{m}\wedge\hat{m}^{*}
+{\textstyle\frac{1}{2}} R_{uu}\hat{l}\wedge \hat{m}
+C\hat{l}\wedge \hat{m}^{*}\, ,
\end{equation}

\noindent
where $C$ is a function to be chosen so as to make this equation (and, hence,
Eq.~(\ref{eq:Deta})) integrable.

Once $\zeta$ has been eliminated, we can solve Eq.~(\ref{eq:FIJ}) of
$\mathcal{F}_{IJ}$ as follows:

\begin{equation}
\label{eq:FIJsolution}
\mathcal{F}_{IJ}
= \frac{8\sqrt{2}}{(\Im{\rm m}\, \tau)^{1/2}}
n^{\mu}(\partial_{\mu}\phi_{[I})\phi_{J]}\, .  
\end{equation}


\subsection{Metric}

At this point we need information about the exact form of the metric.  The
most important piece of information comes from the covariant constancy of the
null vector $l^{\mu}$.  Metrics admitting a covariantly constant null vector
are known as $pp$-wave metrics and were first described by Brinkmann in
Refs.~\cite{kn:Br1,kn:Br2}. Since $l^{\mu}$ is a Killing vector and
$d\hat{l}=0$ we can introduce the coordinates $u$ and $v$

\begin{eqnarray}
l_{\mu}dx^{\mu} & \equiv & du\, , \label{u}\\
& & \nonumber \\
l^{\mu}\partial_{\mu}  & \equiv & \frac{\partial}{\partial v}\, . \label{v}
\end{eqnarray}

\noindent
The previous results imply that all the objects we are dealing with
($\tau,\phi_{I},\mathcal{F}_{IJ}$) are independent of $v$.

Using these coordinates, a 4-dimensional $pp$-wave metric takes the
form\footnote{The components of the connection and the Ricci tensor of this
  metric can be found in Appendix~\ref{sec-brinkmannmetric}.  }

\begin{equation}
\label{eq:conformastat}
ds^{2} = 2 du (dv + K du +\omega)
-2e^{2U}dzdz^{*}\, ,
\hspace{1cm}
\omega=\omega_{\underline{z}}dz +\omega_{\underline{z}^{*}}dz^{*}\, ,
\end{equation}

\noindent 
where all the functions in the metric are independent of $v$ and where either
$K$ or the 1-form $\omega$ could, in principle, be removed by a coordinate
transformation.  In this case, however, we have to be very careful because we
have already used part of the freedom we had to redefine the spinors, and,
therefore, the null tetrad, and we have to check that the tetrad integrability
equations (\ref{eq:dtetrad1})-(\ref{eq:dtetrad3}) are satisfied by our choices
of $e^{U},K$ and $\omega$.

We are now ready to study and solve each case separately.


\subsection{Case $A$}

This is the $B=0$ case. $d\tau = A\hat{l}$ implies that $\tau=\tau(u)$ and
$A=\dot{\tau}$. The connection $Q$ can be integrated

\begin{equation}
Q=d\beta(u)\, ,  
\end{equation}

\noindent
and can be eliminated from all the equations by absorbing a phase into the
spinors:

\begin{equation}
e^{-i\beta}\epsilon =\epsilon^{\prime}\, ,
\hspace{1cm} 
e^{i\beta}\eta =\eta^{\prime}\, ,
\end{equation}

\noindent
and similarly on the null tetrad.

To fix the form of the metric, we study the antisymmetric part of
Eq.~(\ref{eq:dtetrad3})

\begin{equation}
d\hat{m} +\hat{a}\wedge\hat{l}= dU\wedge \hat{m} 
+\hat{a}\wedge\hat{l}=0\, ,
\end{equation}

\noindent
which implies that $U$ only depends on $u$ and

\begin{equation}
\label{eq:aA}
\hat{a}= \dot{U}\hat{m} +C\hat{l}\, ,
\end{equation}

\noindent
where $D$ is a function to be found. Substituting into the antisymmetric part
of Eq.~(\ref{eq:dtetrad2}) we find

\begin{equation}
d\hat{n}+\hat{a}^{*}\wedge \hat{m} +\hat{a}\wedge \hat{m}^{*}=
d\hat{n}+C^{*}\hat{l}\wedge \hat{m} +C\hat{l}\wedge \hat{m}^{*}
=0\, ,
\end{equation}

\noindent
which is solved by

\begin{equation}
n=dv +Kdu\, ,
\hspace{1cm} 
C^{*}=-e^{-U}\partial_{\underline{z}}K\, .
\end{equation}

Now, comparing Eq.~(\ref{eq:aA}) with Eq.~(\ref{eq:da}) we find that
$R_{uz}=0$ which implies (since $\omega=0$) that $\dot{U}=0$.

Finally, to ensure supersymmetry, the integrability conditions
Eqs.~(\ref{eq:ksi9})-(\ref{eq:ksi11}) have to be satisfied, and, with constant
$U$ all of them are  automatically satisfied.

It also follows form the previous equations that the $\phi_{I}$s can only
depend on $u$ and $\mathcal{F}_{IJ}$ is given by

\begin{equation}
\label{eq:FIJsolution2}
\mathcal{F}_{IJ}
= \frac{8\sqrt{2}}{(\Im{\rm m}\, \tau)^{1/2}}
\dot{\phi}_{[I}\phi_{J]}\, .  
\end{equation}

Now, let us consider the equations of motion. The scalar, Maxwell and Bianchi
equations are automatically satisfied and the Einstein equation can be solved
by a $K$ satisfying

\begin{equation}
2\partial_{\underline{z}} \partial_{\underline{z}^{*}}K= 
\frac{|\dot{\tau}|^{2}}{(\Im {\rm m}\, \tau)^{2}} 
+{\textstyle\frac{1}{16}} \Im {\rm m}\, \tau\, \mathcal{F}^{2}\, .
\end{equation}

These solutions preserve generically $1/4$ of the supersymmetries.


\subsection{Case $B$}

This is the $A=0$ case. If we choose $m^{*}=e^{U}dz^{*}$, then $d\tau=Bm^{*}$
implies $\tau=\tau(z^{*})$ and $Be^{U}=\partial_{\underline{z}^{*}}\tau$.
Substituting the corresponding connection 1-form $Q$ into
Eq.~(\ref{eq:dtetrad3}) one finds

\begin{eqnarray}
B^{*} & = & \frac{g(z,u)}{(\Im {\rm m}\,\tau)^{1/2}}\, ,\\
& & \nonumber \\
\hat{a} & = & -\partial_{\underline{u}}\ln{g}\, \hat{m} +D\hat{l}\, ,
\end{eqnarray}

\noindent
where $g$ is a holomorphic function of $z$ and $D$ is a function to be
determined. The first of these relations tells us that

\begin{equation}
\partial_{\underline{z}}\tau^{*}=\frac{e^{U}}{(\Im {\rm m}\,\tau)^{1/2}}
g(z,u)\, ,
\end{equation}

\noindent
is a holomorphic function of $z$, independent of $u$, and taking the
derivative of both sides with respect to $z^{*}$ we get 

\begin{equation}
\frac{e^{U}}{(\Im {\rm m}\,\tau)^{1/2}}=f(u)\, ,
\hspace{1cm}
g(z,u)=\frac{h(z)}{f(u)}\, ,
\end{equation}

\noindent
where $f(u)$ is a real function of $u$.

Substituting now $\hat{a}$ into the antisymmetric part of
Eq.~(\ref{eq:dtetrad2}) we find that $\hat{n}$ is given by

\begin{equation}
\hat{n} = dv +\omega\, ,  
\end{equation}

\noindent
(so $K=0$ in the metric Eq.~(\ref{eq:conformastat})) where the 1-form $\omega$
satisfies

\begin{equation}
\label{eq:omegaequation1}
f_{\underline{z}\underline{z}^{*}}=e^{2U}\partial_{\underline{u}}
\ln{(B/B^{*})}=0\, ,
\end{equation}

\noindent
and $D$ is given by 

\begin{equation}
D^{*}=-\dot{\omega}_{\underline{z}}e^{-U}\, .
\end{equation}

Now that we have determined $\hat{a}$ we have to check that it satisfies the
integrability condition Eq.~(\ref{eq:da}). This requires the following
equations to be satisfied:

\begin{eqnarray}
R_{uz^{*}} +{\textstyle\frac{i}{2}}
\frac{\partial_{\underline{u}}\ln{f}B}{\Im {\rm m}\,\tau} & = & 0\, ,\\
& & \nonumber \\
R_{uu} -[\partial_{\underline{u}}^{2}\ln{f}
+\partial_{\underline{u}}\ln{f}\partial_{\underline{u}}\ln{f}]
-2e^{-U}\partial_{\underline{z}}D & = & 0\, ,\\
& & \nonumber \\
C -e^{-U}\partial_{\underline{z}^{*}}D & = & 0\, .
\end{eqnarray}

Comparing with the integrability conditions
Eqs.~(\ref{eq:ksi9})-(\ref{eq:ksi11}), we conclude that $f$ must be a constant
that we normalize $f=1$ and that $\omega$ must be exact, and we can eliminate
it. Further, the $\phi_{I}$s must be constant and the vector field strengths
must vanish. 

All the equations of motion are automatically satisfied in these conditions,
and the solutions are the \textit{stringy cosmic strings} of
Ref.~\cite{Greene:1989ya}. 

Our result differs from Tod's, who used $\tau$ and $\tau^{*}$ as coordinates
and found very similar solutions with nontrivial $\omega$ that depend in a
very complicated way on a function $g(\tau,u)$ an its complex conjugate. This
function could be eliminated by a coordinate change in which all the $u$
dependence and the 1-form $\omega$ disappear, recovering the stringy cosmic
string solutions.


\section*{Acknowledgements}

The authors would like to thank P.~Meessen for interesting and insightful
conversations and A.~Fern\'andez del R\'{\i}o for her help in the early stages
of this work.  T.O.~is indebted to M.M.~Fern\'andez for her support. This work
has been supported in part by the Spanish grant BFM2003-01090.

\appendix

\section{Conventions}
\label{sec-conventions}


\subsection{Tensors}

We use Greek letters $\mu,\nu,\rho,\ldots$ as (\textit{curved}) tensor indices
in a coordinate basis and Latin letters $a,b,c\ldots$ as (\textit{flat})
tensor indices in a tetrad basis.  Underlined indices are always curved
indices. We symmetrize $()$ and antisymmetrize $[]$ with weight one
(i.e.~dividing by $n!$). We use mostly minus signature $(+---)$. $\eta$ is the
Minkowski metric and a general metric is denoted by $g$. Flat and curved
indices are related by tetrads $e_{a}{}^{\mu}$ and their inverses
$e^{a}{}_{\mu}$, satisfying

\begin{equation}
e_{a}{}^{\mu}e_{b}{}^{\nu}g_{\mu\nu}=\eta_{ab}\, ,
\hspace{1cm}
e^{a}{}_{\mu}e^{b}{}_{\nu}\eta_{ab}=g_{\mu\nu}\, .  
\end{equation}

$\nabla$ is the total (general- and Lorentz-) covariant derivative, whose
action on tensors and spinors ($\psi$) is given by

\begin{equation}
\begin{array}{rcl}
\nabla_{\mu}\xi^{\nu} & =  & \partial_{\mu}\xi^{\nu}
+\Gamma_{\mu\rho}{}^{\nu}\xi^{\rho}\, , \\
& & \\
\nabla_{\mu}\xi^{a}  & =  &  \partial_{\mu}\xi^{a} 
+\omega_{\mu b}{}^{a} \xi^{b}\, , \\
& & \\
\nabla_{\mu}\psi 
& = & \partial_{\mu} \psi -{\textstyle\frac{1}{4}}
\omega_{\mu}{}^{ab}\gamma_{ab}\psi\, ,\\
\end{array}
\end{equation}

\noindent 
where $\gamma_{ab}$ is the antisymmetric product of two gamma matrices (see
next section), $\omega_{\mu b}{}^{a}$ is the spin connection and
$\Gamma_{\mu\rho}{}^{\nu}$ is the affine connection. The respective curvatures
are defined through the Ricci identities

\begin{equation}
  \begin{array}{rcl}
\left[ \nabla_{\mu} , \nabla_{\nu} \right]\ \xi^{\rho} & = &
R_{\mu\nu\sigma}{}^{\rho}(\Gamma)\, \xi^{\sigma} 
+T_{\mu\nu}{}^{\sigma}\nabla_{\sigma}  \xi^{\rho}\, , \\
& & \\
\left[ \nabla_{\mu} , \nabla_{\nu} \right]\, \xi^{a} & = &
R_{\mu\nu b}{}^{a} (\omega)\xi^{b}\, ,\\
& & \\
\left[ \nabla_{\mu} , \nabla_{\nu} \right]\, \psi & = &
-\frac{1}{4}R_{\mu\nu}{}^{ab}(\omega) \gamma_{ab}\psi\, .\\
\end{array}
\end{equation}

\noindent 
and given in terms of the connections by

\begin{equation}
\label{eq:curvatures}
\begin{array}{rcl}
R_{\mu\nu\rho}{}^{\sigma}(\Gamma) & = & 
2\partial_{[\mu}\Gamma_{\nu]\rho}{}^{\sigma}
+ 2\Gamma_{[\mu|\lambda}{}^{\sigma} \Gamma_{\nu]\rho}{}^{\lambda}\, ,\\
& & \\
R_{\mu\nu a}{}^{b} (\omega) & = & 2\partial_{[\mu}\, \omega_{\nu] a}{}^{b} 
-2\omega_{[\mu| a}{}^{c}\,\omega_{|\nu]c}{}^{b}\, .\\
\end{array}  
\end{equation}

These two connections are related by the tetrad postulate

\begin{equation}
\nabla_{\mu}e_{a}{}^{\mu}=0\, ,  
\end{equation}

\noindent
by

\begin{equation}
\omega_{\mu a}{}^{b} = \Gamma_{\mu a}{}^{b} 
+e_{a}{}^{\nu}\partial_{\mu}e_{\nu}{}^{b}\, ,
\end{equation}

\noindent
which implies that the curvatures are, in turn, related by

\begin{equation}
R_{\mu\nu\rho}{}^{\sigma}(\Gamma) = e_{\rho}{}^{a} e^{\sigma}{}_{b}
R_{\mu\nu a}{}^{b}(\omega)\, .
\end{equation}

Finally, metric compatibility and torsionlessness fully determine the
connections to be of the form

\begin{equation}
  \begin{array}{rcl}
\Gamma_{\mu\nu}{}^{\rho} & = &
{\textstyle\frac{1}{2}}g^{\rho\sigma}
\left\{\partial_{\mu}g_{\nu\sigma} +\partial_{\nu}g_{\mu\sigma}
-\partial_{\sigma}g_{\mu\nu} \right\}\, ,\\
& & \\
\omega_{abc} & = & -\Omega_{abc}+\Omega_{bca} -\Omega_{cab}\, ,
\hspace{1cm}
\Omega_{ab}{}^{c} = 
e_{a}{}^{\mu}e_{b}{}^{\nu} \partial_{[\mu}e^{c}{}_{\nu]}\, .\\
\end{array}
\end{equation}


The 4-dimensional fully antisymmetric tensor is defined in flat indices by
tangent space by
\begin{equation}
\epsilon^{0123}= +1\, ,  
\hspace{.5cm}
\Rightarrow
\epsilon_{013} = -1\, ,
\end{equation}

\noindent 
and in curved indices by
\begin{equation}
\epsilon^{\mu_{1}\cdots \mu_{3}} = \sqrt{|g|}\, 
e^{\mu_{1}}{}_{a_{1}} \cdots e^{\mu_{3}}{}_{a_{3}} 
\epsilon^{a_{3}\cdots a_{3}}\, ,
\end{equation}

\noindent
so, with upper indices, is independent of the metric and has the same value as
with flat indices.

We define the (Hodge) dual of a completely antisymmetric tensor of rank $k$,
$F_{(k)}$ by

\begin{equation}
{}^{\star}F_{(k)}{}^{\mu_{1}\cdots \mu_{(d-k)}} = 
{\textstyle\frac{1}{k! \sqrt{|g|}}} 
\epsilon^{\mu_{1}\cdots\mu_{(d-k)}\mu_{(d-k+1)}\cdots\mu_{d}}
F_{(k)\mu_{(d-k+1)}\cdots\mu_{d}}\, .
\end{equation}

Differential forms of rank $k$ are normalized as follows:

\begin{equation}
F_{(k)}\equiv   
{\textstyle\frac{1}{k!}} F_{(k)}{}^{\mu_{1}\cdots \mu_{k}} dx^{1}\wedge
\cdots dx^{k}\, .
\end{equation}

For any 4-dimensional 2-form, we define

\begin{equation}
F^{\pm}\equiv {\textstyle\frac{1}{2}}(F\pm i\,{}^{\star}\!F)\, ,
\hspace{1cm}
\pm i {}^{\star} \!F^{\pm}=F^{\pm}\, .
\end{equation}

For any two 2-forms $F,G$, we have 

\begin{equation}
F^{\pm}\cdot G^{\mp}=0\, ,
\hspace{1cm}
F^{\pm}{}_{[\mu}{}^{\rho}\cdot G^{\mp}{}_{\nu]\rho}=0\, .
\end{equation}

Given any 2-form $F=\frac{1}{2}F_{\mu\nu}dx^{\mu}\wedge dx^{\nu}$ and a
non-null 1-form $\hat{V}=V_{\mu}dx^{\mu}$, we can express $F$ in the form

\begin{equation}
\label{eq:decomposition1}
F=V^{-2}[E\wedge \hat{V} -{}^{\star}\!(B\wedge \hat{V})]\, ,
\hspace{1cm}
E_{\mu}\equiv F_{\mu\nu}V^{\nu}\, ,
\hspace{1cm}
B_{\mu}\equiv {}^{\star}\!\!F_{\mu\nu}V^{\nu}\, .
\end{equation}

For the complex combinations $F^{\pm}$ we have

\begin{equation}
\label{eq:decomposition2}
F^{\pm}=V^{-2}[C^{\pm}\wedge \hat{V} 
\pm i\,{}^{\star}\!(C^{\pm}\wedge \hat{V})]\, ,
\hspace{1cm}
C^{\pm}_{\mu}\equiv F^{\pm}_{\mu\nu}V^{\nu}\, .
\end{equation}

If we have a (real) null vector $l^{\mu}$, we can always add three more
null vectors $n^{\mu},m^{\mu},m^{*\, \mu}$ to construct a complex null tetrad
such that the local metric in this basis takes the form

\begin{equation}
\label{eq:nulltetradmetric}
 \left(\begin{array}{rrrr}
     0 & 1 & 0 & 0 \\
     1 & 0 & 0 & 0 \\
     0 & 0 & 0 & -1 \\
     0 & 0 & -1 & 0 \\
 \end{array}\right)
\end{equation}

\noindent
with the ordering $\left(l,n,m,m^{*}\right)$.  For the local volume element
we obtain $\epsilon^{lnmm^{*}}=i$.  The general expansion in the dual basis
of 1-forms $\left(\hat{l},\hat{n},\hat{m},\hat{m}^{*}\right)$ of $F^{+}$
depends on three arbitrary complex functions $a,b,c$

\begin{equation}
 F^{+}  =  a\left(\hat{l}\wedge \hat{n}+\hat{m}\wedge\hat{m}^{*}\right)
+b\hat{l}\wedge \hat{m}^{*} +c \hat{n}\wedge \hat{m}\, ,
\hspace{1cm}
F^{-}=(F^{+})^{*}\, .
\end{equation}

\noindent
Then, in this case, $F$ is not completely determined by its contraction with
the null vector $l$, but

\begin{equation}
\label{eq:decomposition3}
F^{+}  =  L^{\pm}\wedge \hat{n} \pm {}^{\star}( L^{\pm}\wedge \hat{n})
+b\hat{l}\wedge \hat{m}\, ,
\hspace{1cm}
L^{\pm}_{\mu}\equiv F^{\pm}{}_{\mu\nu}l^{\nu}= al_{\mu} -c m_{\mu}\, .
\end{equation}


\subsection{Gamma matrices and spinors}

We work with a purely imaginary representation

\begin{equation}
\gamma^{a\, *}= -\gamma^{a}\, ,  
\end{equation}

\noindent 
and our convention for their anticommutator is

\begin{equation}
\{\gamma^{a},\gamma^{b}\}= +2\eta^{ab}\, .  
\end{equation}

Thus, 

\begin{equation}
\gamma^{0}\gamma^{a}\gamma^{0}=
\gamma^{a\, \dagger}= \gamma^{a\, -1}=\gamma_{a}\, .
\end{equation}

The chirality matrix is defined by

\begin{equation}
\gamma_{5}\equiv -i\gamma^{0}\gamma^{1}\gamma^{2}\gamma^{3}
={\textstyle\frac{i}{4!}} \epsilon_{abcd}
\gamma^{a}\gamma^{b}\gamma^{c}\gamma^{d}\, ,
\end{equation}

\noindent
and satisfies

\begin{equation}
\gamma_{5}{}^{\dagger}=-\gamma_{5}{}^{*}=\gamma_{5}\, ,
\hspace{1cm}
(\gamma_{5})^{2}=1\, .
\end{equation}

With this chirality matrix, we have the identity

\begin{equation}
\label{eq:dualgammaidentityind4}
\gamma^{a_{1}\cdots a_{n}} =\frac{(-1)^{\left[n/2\right]}i}{(4-n)!}
\epsilon^{a_{1}\cdots a_{n}b_{1}\cdots b_{4-n}} \gamma_{b_{1}\cdots b_{4-n}}
\gamma_{5}\, .
\end{equation}

Our convention for Dirac conjugation is

\begin{equation}
\bar{\psi}=i\psi^{\dagger}\gamma_{0}\, .  
\end{equation}

Using the identity Eq.~(\ref{eq:dualgammaidentityind4}) the general
$d=4$ Fierz identity for \textit{commuting} spinors takes the form

\begin{equation}
\label{eq:Fierzidentities}
  \begin{array}{rcl}
(\bar{\lambda} M\chi) (\bar{\psi} N \varphi ) & = &
{\textstyle\frac{1}{4}} (\bar{\lambda} M N \varphi) (\bar{\psi} \chi )
+{\textstyle\frac{1}{4}} (\bar{\lambda} M \gamma^{a}N \varphi) 
(\bar{\psi} \gamma_{a}\chi ) 
-{\textstyle\frac{1}{8}} (\bar{\lambda} M \gamma^{ab}N \varphi) 
(\bar{\psi} \gamma_{ab}\chi )
\\
& & \\
& & 
-{\textstyle\frac{1}{4}} (\bar{\lambda} M \gamma^{a}\gamma_{5}N \varphi) 
(\bar{\psi} \gamma_{a}\gamma_{5}\chi )
+{\textstyle\frac{1}{4}} (\bar{\lambda} M \gamma_{5}N \varphi) 
(\bar{\psi}\gamma_{5}\chi )\, .\\
\end{array}
\end{equation}

We use 4-component chiral spinors whose chirality is related to the
position of the $SU(4)$ index:

\begin{equation}
\gamma_{5}\chi_{I}=+ \chi_{I}\, ,
\hspace{1cm}
\gamma_{5}\psi_{\mu\, I}=- \psi_{\mu\, I}\, ,
\hspace{1cm}
\gamma_{5}\epsilon_{I}=- \epsilon_{I}\, .
\end{equation}

\noindent
Both (chirality and position of the $SU(4)$ index) are reversed under
complex conjugation:

\begin{equation}
\gamma_{5}\chi_{I}^{*} \equiv\gamma_{5}\chi^{I}= -\chi^{I}\, ,
\hspace{1cm}
\gamma_{5}\psi_{\mu\, I}^{*}\equiv
\gamma_{5}\psi_{\mu}{}^{I}=+\psi_{\mu}{}^{I}\, ,
\hspace{1cm}
\gamma_{5}\epsilon_{I}^{*}\equiv
\gamma_{5}\epsilon^{I}= +\epsilon^{I}\, .
\end{equation}

We take this fact into account when Dirac-conjugating chiral spinors:

\begin{equation}
\bar{\chi}^{I}\equiv i(\chi_{I})^{\dagger}\gamma_{0}\, ,
\hspace{.5cm}
\bar{\chi}^{I}\gamma_{5}=-\bar{\chi}^{I}\, ,\,\,\,\, {\rm etc.}
\end{equation}

The sum of the two chiral spinors related by complex conjugation gives
a standard (real) Majorana spinor with an $SU(4)$ index with the
complicated transformation rule of Ref.~\cite{Cremmer:1977tt}.


\section{Fierz identities for bilinears}
\label{sec-Fierz}

Here we are going to work with an arbitrary number $N$ of chiral
spinors, although we are ultimately interested in the $N=4$ case only.
Whenever there are special results for particular values of $N$, we
will explicitly say so. We should bear in mind that the maximal number
of independent chiral spinors is 2 and, for $N>2$ (in particular for
$N=4$) $N$ spinors cannot be linearly independent at a given point.
This trivial fact has important consequences.

Given $N$ chiral commuting spinors $\epsilon_{I}$ and their complex
conjugates $\epsilon^{I}$ we can constructed the following bilinears
that are not obviously related via
Eq.~(\ref{eq:dualgammaidentityind4}):

\begin{enumerate}
\item A complex matrix of scalars

\begin{equation}
M_{IJ}\equiv \bar{\epsilon}_{I}\epsilon_{J}\, ,  
\hspace{1cm}
M^{IJ}\equiv \bar{\epsilon}^{I}\epsilon^{J}=(M_{IJ})^{*}\, ,
\end{equation}

\noindent
which is antisymmetric $M_{IJ}=-M_{JI}$.

\item A complex matrix of vectors

\begin{equation}
V^{I}{}_{J\, a}\equiv i\bar{\epsilon}^{I}\gamma_{a}\epsilon_{J}\, ,  
\hspace{1cm}
V_{I}{}^{J}{}_{a}\equiv i\bar{\epsilon}_{I}\gamma_{a}\epsilon^{J}
=(V^{I}{}_{J\, a})^{*}\, ,
\end{equation}

\noindent
which is Hermitean:

\begin{equation}
(V^{I}{}_{J\, a})^{*}=V_{I}{}^{J}{}_{a} = V^{J}{}_{I\, a}
=(V^{I}{}_{J\, a})^{T}\, .  
\end{equation}

\item A complex matrix of 2-forms

\begin{equation}
\label{eq:p}
\Phi_{IJ\, ab}\equiv \bar{\epsilon}_{I}\gamma_{ab}\epsilon_{J}\, ,  
\hspace{1cm}
\Phi^{IJ}{}_{ab}\equiv 
\bar{\epsilon}^{I}\gamma_{ab}\epsilon^{J}=(M_{IJ})^{*}\, ,
\end{equation}

\noindent
which is symmetric in the $SU(N)$ indices $\Phi_{IJ\, ab}=\Phi_{JI\,
  ab}$ and, further,

\begin{equation}
\label{eq:p1}
{}^{\star}\!\Phi_{IJ\, ab}=-i\Phi_{IJ\, ab}\,\,\, \Rightarrow\,\,\,
\Phi_{IJ\, ab} = \Phi_{IJ}{}^{+}{}_{ab}\, .
\end{equation}

\noindent
As we are going to see, this matrix of 2-forms can be expressed
entirely in terms of the scalar and vector bilinears.

\end{enumerate}

It is straightforward to get identities for the products of these
bilinears using the Fierz identity Eq.~(\ref{eq:Fierzidentities}).
First, the products of scalars:

\begin{eqnarray}
M_{IJ}M_{KL} & = & {\textstyle\frac{1}{2}}M_{IL}M_{KJ} 
- {\textstyle\frac{1}{8}}\Phi_{IL}\cdot\Phi_{KJ}\, ,\label{eq:mm1}\\
& & \nonumber \\
M_{IJ}M^{KL} & = & -{\textstyle\frac{1}{2}} V^{L}{}_{I}\cdot V^{K}{}_{J}
\label{eq:mm2}\, .
\end{eqnarray}

\noindent
From Eq.~(\ref{eq:mm1}) immediately follows

\begin{equation}
\label{eq:mm3}
M_{I[J}M_{KL]}=0\, ,  
\end{equation}

\noindent
which is a particular case of the Fierz identity

\begin{equation}
\label{eq:mm4}
\epsilon_{[J}M_{KL]}=0\, .  
\end{equation}

\noindent
For $N=4,8,\ldots$, Eq.~(\ref{eq:mm3}) implies, in turn

\begin{equation}
{\rm Pf}\, M=0\,\,\, \Rightarrow {\rm det}\, M =0\, .  
\end{equation}

\noindent
For $N=4$ we can define the $SU(4)$-dual of $M_{IJ}$

\begin{equation}
\tilde{M}_{IJ}\equiv {\textstyle\frac{1}{2}}\varepsilon_{IJKL}M^{KL}\, ,  
\hspace{1cm}
\varepsilon^{1234}=\varepsilon_{1234}=+1\, ,
\end{equation}

\noindent
and the vanishing of the Pfaffian implies

\begin{equation}
\tilde{M}_{IJ} M^{IJ}=0\, .  
\end{equation}

\noindent
From Eq.~(\ref{eq:mm2}) and the antisymmetry of $M$ immediately
follows

\begin{equation}
\label{eq:vv1}
V^{I}{}_{L}\cdot V^{K}{}_{J}= -V^{I}{}_{J}\cdot V^{K}{}_{L}
=-V^{K}{}_{L}\cdot V^{I}{}_{J}\, ,
\end{equation}

\noindent
which implies that all the vector bilinears $V^{I}{}_{J\, a}$ are null:

\begin{equation}
\label{eq:vv2}
V^{I}{}_{J}\cdot V^{I}{}_{J}=0\, .  
\end{equation}

\noindent
On the other hand, from Eqs.~(\ref{eq:vv1}) and (\ref{eq:mm2})
follows the real $SU(N)$-invariant combination of vectors $V_{a}\equiv
V^{I}{}_{I\, a}$ is always non-spacelike:

\begin{equation}
\label{eq:vv3}
V^{2}=-V^{I}{}_{J}\cdot V^{J}{}_{I} = 2 M^{IJ}M_{IJ}\geq 0\, .
\end{equation}

The products of $M$ with the other bilinears\footnote{We omit the
  product $M_{IJ}\Phi_{KL\, ab}$ which will not be used.} give

\begin{eqnarray}
M_{IJ}V^{K}{}_{L\, a} & = & {\textstyle\frac{1}{2}}M_{IL}V^{K}{}_{J\, a} 
+{\textstyle\frac{1}{2}}\Phi_{IL\, ba} V^{K}{}_{J}{}^{b}\, ,
\label{eq:mv1}\\
& & \nonumber \\
M_{IJ}\Phi^{KL}{}_{ab} & = & 
V^{L}{}_{I\, [a|} V^{K}{}_{J\, |b]} 
-{\textstyle\frac{i}{2}} \epsilon_{ab}{}^{cd}
V^{L}{}_{I\, c} V^{K}{}_{J\, d} 
\label{eq:mp1}\, .
\end{eqnarray}

Now, let us consider the product of two arbitrary vectors\footnote{The product
  $V^{I}{}_{J\, a} V_{L}{}^{K}{}_{b}$ gives a different identity that will not
  be used}:

\begin{equation}
\label{eq:vv4}
V^{I}{}_{J\, a} V^{K}{}_{L\, b}
= {\textstyle\frac{i}{2}} \epsilon_{ab}{}^{cd}
V^{I}{}_{L\, c} V^{K}{}_{J\, d} 
+V^{I}{}_{L\, (a|} V^{K}{}_{J\, |b)}
-{\textstyle\frac{1}{2}} g_{ab}
V^{I}{}_{L}\cdot V^{K}{}_{J}\, .
\end{equation}

\noindent
For $V^{2}$ this identity allows us to write the metric in the form

\begin{equation}
\label{eq:vv5}
g_{ab}=2V^{-2}[V_{a}V_{b}-V^{I}{}_{J\, a} V^{J}{}_{I\, b}]\, .
\end{equation}

Following Tod \cite{Tod:1995jf}, for $V^{2}\neq 0$ we  introduce

\begin{equation}
\label{eq:J}
\mathcal{J}^{I}{}_{J} \equiv \frac{2 M^{IK} M_{JK}}{|M|^{2}}
= \frac{2 V\cdot V^{I}{}_{J}}{V^{2}}\, ,
\hspace{1cm}
|M|^{2}\equiv M^{LM}M_{LM}={\textstyle\frac{1}{2}}V^{2}\, .
\end{equation}

\noindent
Using Eq.~(\ref{eq:mm1}) we can show that it is a Hermitean projector
whose trace equals 2:

\begin{equation}
\label{eq:J1}
\mathcal{J}^{I}{}_{J}\mathcal{J}^{J}{}_{K}=
\mathcal{J}^{I}{}_{K}\, ,
\hspace{1cm}
\mathcal{J}^{I}{}_{I}=+2\, .
\end{equation}

\noindent
Further, using the general Fierz identity we find 

\begin{equation}
\label{eq:J2}
\mathcal{J}^{I}{}_{J}\epsilon^{J} =\epsilon^{I}\, ,  
\hspace{1cm}
\epsilon_{I}\mathcal{J}^{I}{}_{J} =\epsilon_{J}\, ,  
\end{equation}

\noindent
which should be understood for $N>2$ of the fact that the $\epsilon^{I}$ are
not linearly independent\footnote{For $N=2$
  $\mathcal{J}^{I}{}_{J}=\delta^{I}{}_{J}$. See later on.}. As a consequence
of the above identity, the contraction of $\mathcal{J}$ with any of the
bilinears is the identity.  Using this result and Eq.~(\ref{eq:mp1}), we find

\begin{equation}
\label{eq:p2}
\Phi^{KL}{}_{ab}=\frac{2 M^{IK}M_{IJ}}{|M|^{2}}\Phi^{JL}{}_{ab}=  
\frac{2 M^{IK}}{|M|^{2}}V^{L}{}_{I\, [a}V_{b]} 
-i\frac{M^{IK}}{|M|^{2}}\epsilon_{ab}{}^{cd}V^{L}{}_{I\, c}V_{d}\, .
\end{equation}

Other useful identities are

\begin{equation}
\label{eq:MMJJ}
\frac{M_{IJ}M^{KL}}{|M|^{2}} =
\mathcal{J}^{K}{}_{[I}\mathcal{J}^{L}{}_{J]}\, ,
\end{equation}

\noindent
and

\begin{equation}
 \frac{2 \tilde{M}^{IK}
 \tilde{M}_{JK}}{|M|^{2}}=\delta^{I}{}_{J}-\mathcal{J}^{I}{}_{J}
 \equiv \tilde{\mathcal{J}}{}^{I}{}_{J}\, ,
\end{equation}

\noindent
which is the complementary projector.

In the null case $V^{2}=|M|^{2}=0$ it is customary to write $l_{a}\equiv
V^{I}{}_{I\, a}$.  Since $|M|^{2}$ is a sum of positive numbers, each of them
must vanish independently, i.e.~$M^{IJ}=0$. This implies that all spinors
$\epsilon^{I}$ are proportional and one can write 

\begin{equation}
\label{proportional}
  \epsilon_{I}=\phi_{I} \epsilon\, ,
\end{equation}

\noindent
for some complex functions $\phi_{I}$ which transform as an $SU(4)$ vector,
and some negative-chirality spinor $\epsilon$. These are defined up to a
rescaling by a complex function and opposite weights. Part of this freedom can
be fixed by normalizing

\begin{equation}
\label{eq:normalization}
  \phi_{I}\phi^{I}=1\, ,
\hspace{1cm}
\phi^{I}\equiv\phi_{I}^{*}\, .
\end{equation}

Then, the only freedom that remains in the definition of $\phi^{I}$ is a
change by a local phase $\theta(x)$

\begin{equation}
\label{freephase} 
\phi_{I}\rightarrow  e^{i\theta}\phi_{I}\, ,
\hspace{1cm}
\epsilon\rightarrow e^{-i\theta}\epsilon\, .
\end{equation}

In this case on can construct another Hermitean projector
$\mathcal{K}^{I}{}_{J}$ that plays a role analogous to that of
$\mathcal{J}^{I}{}_{J}$ in the non-null case:

\begin{equation}
\mathcal{K}^{I}{}_{J} \equiv \phi^{I}\phi_{J}\, ,
\end{equation}

\noindent
which satisfies

\begin{equation}
\label{eq:K1}
\mathcal{K}^{I}{}_{J}\mathcal{K}^{J}{}_{K}=
\mathcal{K}^{I}{}_{K}\, ,
\hspace{1cm}
\mathcal{K}^{I}{}_{I}=+1\, ,
\end{equation}

\noindent
and

\begin{equation}
\label{eq:K2}
\mathcal{K}^{I}{}_{J}\epsilon^{J} =\epsilon^{I}\, ,  
\hspace{1cm}
\epsilon_{I}\mathcal{K}^{I}{}_{J} =\epsilon_{J}\, ,  
\end{equation}

\noindent
which expresses the known fact that only one spinor is linearly independent in
this case.

In the null case, all the vector bilinears are also proportional to the null
vector $l$:

\begin{equation}
V^{I}{}_{J\, a} = \mathcal{K}^{I}{}_{J}l_{a}\, .  
\end{equation}

Once $\epsilon$ is given, we may introduce an auxiliary spinor with the same
chirality and opposite $U(1)$ charge as $\epsilon$ and normalized against
$\epsilon$ by

\begin{equation}
\label{eq:auxiliary}
  \bar{\epsilon}\eta=\frac{1}{2}\, ,
\end{equation}

\noindent
where $\bar{\epsilon}=i\epsilon^{T}\gamma_{0}$.   With both spinors we can
construct a complex null tetrad with metric Eq.~(\ref{eq:nulltetradmetric}) as
follows:

\begin{equation}
\label{eq:nulltetraddef}
l_{\mu}=i\bar{\epsilon^{*}}\gamma_{\mu}\epsilon\, ,
\hspace{.5cm}
n_{\mu}=i\bar{\eta^{*}}\gamma_{\mu}\eta\, ,
\hspace{.5cm}
m_{\mu}=i\bar{\epsilon^{*}}\gamma_{\mu}\eta=
i\bar{\eta}\gamma_{\mu}\epsilon^{*}\, ,
\hspace{.5cm}
m_{\mu}^{*}=i\bar{\epsilon}\gamma_{\mu}\eta^{*}=
i\bar{\eta^{*}}\gamma_{\mu}\epsilon\, .
\end{equation}

The normalization condition (\ref{eq:normalization}) does not fix completely
the auxiliary spinor $\eta$ and the freedom in the choice of $\eta$ becomes a
freedom in the null tetrad. First of all, there is a $U(1)$ freedom
Eq.~(\ref{freephase}) under which $\eta^{\prime}=e^{i\theta}\eta$ and

\begin{equation}
l^{\prime} = l\, ,
\hspace{.5cm}
n^{\prime} = n\, ,
\hspace{.5cm}
m^{\prime} = e^{2i\theta} m\, .
\end{equation}

Further, we can also shift $\eta$ by terms proportional to $\epsilon$
preserving the normalization

\begin{equation}
\label{eq:redef}
\eta^{\prime}=\eta +\delta \epsilon\, .  
\end{equation}

\noindent
Under this redefinition of $\eta$, the null tetrad transforms as follows:

\begin{equation}
\label{eq:redef2}
l^{\prime}=l\, ,
\hspace{.5cm}
n^{\prime} = n+\delta^{*}m+ \delta m^{*} +|\delta|^{2}l\, ,
\hspace{.5cm}
m^{\prime} = m +\delta l\, .
\end{equation}






\subsection{The $N=2$ case}

Here we describe some of the peculiarities of the $N=2$ case in which the
number of spinors is precisely the necessary to construct a basis at each
point.

In the $N=2$ case there is only one independent (complex) scalar $X$ since

\begin{equation}
\bar{\epsilon}_{I}\epsilon_{J}=X\epsilon_{IJ}\, ,  
\end{equation}

\noindent
where $\epsilon_{IJ}$ is the (constant) 2-dimensional totally antisymmetric
tensor. It follows that

\begin{equation}
|M|^{2}=2|X|^{2}\, ,  
\end{equation}

\noindent
and, using $\epsilon_{IJ}\epsilon^{KL}=\delta_{IJ}{}^{KL}$
we can show that the projector

\begin{equation}
\mathcal{J}^{I}{}_{J}=\delta^{I}{}_{J}\, .  
\end{equation}

In the $|M|^{2}\neq 0$ case, the four vector bilinears $V^{I}{}_{J\, \mu}$ can
be used as a null tetrad

\begin{equation}
l_{\mu}= V^{1}{}_{1\, \mu}\, ,\,\,\,\,
n_{\mu}= V^{2}{}_{2\, \mu}\, ,\,\,\,\,
m_{\mu}= V^{1}{}_{2\, \mu}\, ,\,\,\,\,
m^{*}_{\mu}= V^{2}{}_{1\, \mu}\, ,.
\end{equation}

\noindent
Alternatively. one can use the four combinations

\begin{equation}
V^{a}{}_{\mu}\equiv  {\textstyle\frac{1}{\sqrt{2}}}
V^{I}{}_{J\, \mu}(\sigma^{a})^{J}{}_{I}\, ,
\end{equation}

\noindent
with $\sigma^{0}=1$ and $\sigma^{i}$ the three (traceless, Hermitean) Pauli
matrices as an orthonormal tetrad in which $V^{0}$ is timelike and the $V^{i}$
are spacelike.


\section{Connection and curvature of the conformastationary metric}
\label{sec-conformastationarymetric}

A conformastationary metric has the general form

\begin{equation}
ds^{2} = |M|^{2}(dt+\omega)^{2} 
-|M|^{-2}\gamma_{\underline{i}\underline{j}}dx^{i}dx^{j}\, ,
\hspace{1cm}
i,j=1,2,3\, ,
\end{equation}

\noindent
where all components of the metric are independent of the time coordinate $t$.
Choosing the Vielbein basis

\begin{equation}
(e_{\mu}{}^{a}) = 
\left(
  \begin{array}{cc}
|M| & |M| \omega_{\underline{i}} \\
& \\
0 & |M|^{-1} v_{\underline{i}}{}^{j} \\
  \end{array}
\right)\, ,
\hspace{1cm}
(e_{a}{}^{\mu}) = 
\left(
  \begin{array}{cc}
|M|^{-1} & -|M| \omega_{i} \\
& \\
0 & |M| v_{i}{}^{\underline{j}} \\
  \end{array}
\right)\, ,
\end{equation}

\noindent
where 

\begin{equation}
\gamma_{\underline{i}\underline{j}}
=v_{\underline{i}}{}^{k}v_{\underline{j}}{}^{l}\delta_{kl}\, ,
\hspace{1cm}
v_{i}{}^{\underline{k}}v_{\underline{k}}{}^{j}v_{j}\, ,
\hspace{1cm}
\omega_{i}= v_{i}{}^{\underline{j}}\omega_{\underline{j}}\, ,   
\end{equation}

\noindent
we find that the spin connection components are

\begin{equation}
  \begin{array}{rclrcl}
\omega_{00i} & = & -\partial_{i}|M|\, , \hspace{2cm} &
\omega_{0ij} & = &  \frac{1}{2}f_{ij}\, ,\\
& & & & & \\
\omega_{i0j} & = & \omega_{0ij}\, , &
\omega_{ijk} & = & - |M| o_{ijk} -2 \delta_{i[j}\partial_{k]}|M|\, ,\\
\end{array}
\end{equation}

\noindent
where $o_{i}{}^{jk}$ is the 3-dimensional spin connection and 

\begin{equation}
\partial_{i} \equiv v_{i}{}^{\underline{j}}\partial_{\underline{j}}\, ,
\hspace{1cm}
f_{ij}=  v_{i}{}^{\underline{k}}  v_{j}{}^{\underline{l}}
f_{\underline{k}\underline{l}}\, ,  
\hspace{1cm}
f_{\underline{i}\underline{j}} \equiv
2\partial_{[\underline{i}}\omega_{\underline{j}]}\, .
\end{equation}

The components of the Riemann tensor are 

\begin{equation}
  \begin{array}{rcl}
R_{0i0j} & = & {\textstyle\frac{1}{2}}\nabla_{i}\partial_{j}|M|^{2}
+\partial_{i}|M|\partial_{j}|M| -\delta_{ij}(\partial|M|)^{2}
+{\textstyle\frac{1}{4}}\nabla{i}|M|^{6}f_{ik}f_{jk}\, , \\
& & \\
R_{0ijk} & = & -{\textstyle\frac{1}{2}}\nabla_{i}(|M|^{4}f_{jk})
+{\textstyle\frac{1}{2}}f_{i[j}\partial_{k]}|M|^{4}
-{\textstyle\frac{1}{4}}\delta_{i[j}f_{k]l}\partial_{l}|M|^{4}\, ,\\
& & \\
R_{ijkl} & = & -|M|^{2}R_{ijkl} 
+{\textstyle\frac{1}{2}}|M|^{6}(f_{ij}f_{kl}-f_{k[i}f_{j]l})
-2\delta_{ij,kl}(\partial|M|)^{2}
+4|M|\delta_{[i}{}^{[k}\nabla_{j]}\partial^{l]}|M|\, ,\\
  \end{array}
\end{equation}

\noindent
where all the objects in the right-hand sides of the equations are referred to
the 3-dimensional spatial metric.  The components of the Ricci tensor are

\begin{equation}
  \begin{array}{rcl}
R_{00} & = & -|M|^{2} \nabla^{2}\log{|M|} 
-{\textstyle\frac{1}{4}}|M|^{6}f^{2}\, ,\\
& & \\
R_{0i} & = &  {\textstyle\frac{1}{2}} \nabla_{j}(|M|^{4}f_{ji})\, ,\\
& & \\
R_{ij} & = & |M|^{2}
\{
R_{ij} +2\partial_{i}\log{|M|}\partial_{j}\log{|M|}
-\delta_{ij}\nabla^{2}\log{|M|}
-{\textstyle\frac{1}{2}}|M|^{4}f_{ik}f_{jk}
\}\, ,\\
  \end{array}
\end{equation}

\noindent
and the Ricci scalar is

\begin{equation}
R=-|M|^{2}
\{
R -{\textstyle\frac{1}{4}}|M|^{4}f^{2} -2 \nabla^{2}\log{|M|}
+2(\partial\log{|M|})^{2}
\}\, ,  
\end{equation}


\section{Connection and curvature of a Brinkmann $pp$-wave metric}
\label{sec-brinkmannmetric}

We rewrite here for convenience the 4-dimensional form of these metrics:

\begin{equation}
ds^{2} = 2 du (dv + K du +\omega)
-2e^{2U}dzdz^{*}\, ,
\hspace{1cm}
\omega=\omega_{\underline{z}}dz +\omega_{\underline{z}^{*}}dz^{*}\, ,
\end{equation}

\noindent 
where all the functions in the metric are independent of $v$.

Using also light-cone coordinates in tangent space, a natural Vielbein
basis is

\begin{equation}
\begin{array}{rclclrclcl}
e^{u} & = & du & = & \hat{l}\, ,   
& 
e_{u} & = & \partial_{\underline{u}}-K\partial_{\underline{v}} & = & 
n^{\mu}\partial_{\mu}\, ,  
\\
& & & & & & & & & \\
e^{v} & = & dv+Kdu+\omega & = & \hat{n}\, ,
& 
e_{v} & = & \partial_{\underline{v}} & = & l^{\mu}\partial_{\mu}\, , 
\\
& & & & & & & & & \\
e^{z} & = & e^{U}dz & = & \hat{m} \, , 
& 
e_{z} & = & e^{-U}(\partial_{\underline{z}} 
-\omega_{\underline{z}}\partial_{\underline{v}}) & = & 
-m^{*\, \mu}\partial_{\mu}\, ,\\
& & & & & & & & \\
e^{z^{*}} & = & e^{U}dz^{*} & = & \hat{m}^{*} \, , \hspace{2cm}
& 
e_{z^{*}} & = & e^{-U}(\partial_{\underline{z}^{*}} 
-\omega_{\underline{z}^{*}}\partial_{\underline{v}}) & = & 
-m^{\mu}\partial_{\mu}\, .\\
\end{array}
\end{equation}

\noindent 
The components of the spin connection are

\begin{equation}
\begin{array}{rclrcl}
\omega_{uzu} & = & e^{-U} (\partial_{\underline{z}}K
-\dot{\omega}_{\underline{z}})\, , \hspace{1cm}&
\omega_{uzz^{*}} & = & \textstyle{\frac{1}{2}} e^{-2U}
f_{\underline{z}\underline{z}^{*}} -\dot{U}\, ,\\
 & & & & & \\
\omega_{zz^{*}u} & = & -\textstyle{\frac{1}{2}} e^{-2U}
f_{\underline{z}\underline{z}^{*}} -\dot{U}\, ,  &
\omega_{zzz^{*}} & = & -e^{-U} \partial_{\underline{z}}U\, ,\\
\end{array}
\end{equation}

\noindent 
where $f_{\underline{z}\underline{z}^{*}}=2\partial_{[\underline{z}}
\omega_{\underline{z}^{*}]}$ and a dot stands for partial derivation with
respect to $u$.

The components of the Ricci tensor are

\begin{equation}
\begin{array}{rcl}
R_{zz^{*}} & = & 2e^{-2U} 
\partial_{\underline{z}} \partial_{\underline{z}^{*}} U\, ,\\
& & \\
R_{zu} & = &  
{\textstyle\frac{1}{2}} e^{-3U}\partial_{\underline{z}}
f_{\underline{z}\underline{z}^{*}}
+e^{-U}(\partial_{\underline{z}}\dot{U} 
+\dot{U} \partial_{\underline{z}}U)\, ,\\
& & \\
R_{uu} & = &
-2e^{-2U}\partial_{\underline{z}} \partial_{\underline{z}^{*}} K
+{\textstyle\frac{1}{2}}(f_{\underline{z}\underline{z}^{*}})^{2}
+e^{-2U}(\partial_{\underline{z}}\dot{\omega}_{\underline{z}^{*}}
+\partial_{\underline{z}^{*}}\dot{\omega}_{\underline{z}})
+2(\ddot{U}+\dot{U}\dot{U})\, ,\\
\end{array}
\end{equation}

\noindent 
and the Ricci scalar is just 

\begin{equation}
R=  -4e^{-2U} 
\partial_{\underline{z}} \partial_{\underline{z}^{*}} U\, . 
\end{equation}


\end{document}